\documentclass{aa}  

\usepackage{color}
\usepackage{graphicx}
\usepackage{subcaption}
\usepackage{txfonts}
\usepackage{natbib}
\usepackage{url}
\usepackage{multirow}
\usepackage{xspace}
\bibpunct{(}{)}{;}{a}{}{,}

\newcommand{\teff}{$T_{\rm eff}$}
\newcommand{\nhe} {$N$({\rm He})/$N$({\rm H})}
\newcommand{\msun}{$M_{\rm \odot}$}

\newcommand{\vsini}{$v_{\rm rot}$~sin~$i$}
\newcommand{\agk}{\object{AGK$+$81$\degr$266}}
\newcommand{\lsii}{\object{LS II$+$18$\degr$9}}
\newcommand{\bd}{\object{BD$+$28$\degr$4211}}
\newcommand{\feige}{\object{Feige\,34}}
\newcommand{\feiget}{\object{Feige\,67}}
\usepackage{lineno}
\usepackage{ulem}

\begin{document}

   \title{Spectral analysis of four surprisingly similar hot hydrogen-rich 
subdwarf O stars}

   \author{M. Latour\inst{1}, P. Chayer\inst{2}, 
           E. M. Green\inst{3},  A. Irrgang\inst{1}, and 
           G. Fontaine\inst{4}
          }

   \institute{ Dr. Karl Remeis-Observatory \& ECAP, Astronomical Institute,
               Friedrich-Alexander University Erlangen-N\"{u}rnberg, 
Sternwartstr. 7, 96049, Bamberg, Germany, \email{marilyn.latour@fau.de}
         \and Space Telescope Science Institute, 3700 San Martin Drive, 
Baltimore, MD 21218, USA
         \and Steward Observatory, University of Arizona, 933 North
   Cherry Avenue, Tucson, AZ 85721
         \and D\'epartement de Physique, Universit\'e de Montr\'eal, Succ. 
Centre-Ville, C.P. 6128, Montr\'eal, QC H3C 3J7, Canada
}

   \date{Received 4 July 2017 ; accepted 12 October 2017}

  \abstract
    {Post-extreme horizontal branch stars (post-EHB) are helium-shell burning 
objects evolving away from the EHB and contracting directly towards the white 
dwarf regime. While the stars forming the EHB have been extensively studied in 
the past, their hotter and more evolved progeny are not so well characterized. }
   {We perform a comprehensive spectroscopic analysis of four such bright sdO 
stars, namely \feige, \feiget, \agk, and \lsii, among which the first three are 
used as standard stars for flux calibration. Our goal is to determine their 
atmospheric parameters, chemical properties, and evolutionary status to 
better understand this class of stars that are en route to become white dwarfs. 
}
   {We used non-local thermodynamic equilibrium model atmospheres in combination 
with high quality optical and UV spectra. Photometric data were also used to 
compute the spectroscopic distances of our stars and to characterize the 
companion responsible for the infrared excess of \feige.}
   {The four bright
sdO 
stars have very similar atmospheric parameters with \teff\ 
between 60\,000 and 63\,000 K and log $g$ (cm s$^{-2}$) in the range 5.9 to 
6.1. This places these objects right on the theoretical post-EHB evolutionary tracks. 
The UV spectra are dominated by strong iron and nickel lines and suggest 
abundances that are enriched with respect to those of the Sun by factors of 25 
and 60. On the other hand, the lighter elements, C, N, O, Mg, Si, P, and S, are 
depleted. The stars have very similar abundances, although \agk\ shows 
differences in its light element abundances. For instance, the helium abundance of this object 
is 10 times lower than that observed in the other three stars. All our stars 
show UV spectral lines that require additional line broadening that is 
consistent with a rotational velocity of about 25 km s$^{-1}$. The infrared 
excess of \feige\ is well reproduced by a M0 main-sequence companion and the surface area ratio of the two stars suggests that the system is a physical binary. However, the 
lack of radial velocity variations points towards a low inclination and/or long 
orbital period. Spectroscopic and Hipparcos distances are in good agreement for 
our three brightest stars. }
   {We performed a spectroscopic analysis of four hot sdO stars that are very 
similar in terms of atmospheric parameters and chemical compositions.
  The rotation velocities of our stars are significantly higher than what is 
observed in their immediate progenitors on the EHB, suggesting that angular 
momentum may be conserved as the stars evolve away from the EHB.
}

    
\titlerunning{Spectral analysis of four surprisingly similar hot hydrogen-rich 
subdwarf O stars}             
\authorrunning{M. Latour et al.}

   \maketitle
%

\section{Introduction}


Hot subluminous stars of spectral type B and O (sdB and sdO) form the blue 
extension of the horizontal branch (EHB) in the Hertzsprung-Russell diagram. 
These stars represent the helium burning stage of low mass main-sequence stars 
($\approx$1.0 \msun) that lost almost all of their hydrogen envelope during the 
red giant phase \citep{heb09,heb16}.
The sdO spectral type includes stars that have effective temperatures above 
$\approx$ 38 000 K and surface gravities in the range 
4.6~$\la$~log~$g$~$\la$~6.4. As for their helium abundance, it varies from one 
hundredth to a thousand times solar. Given this large spread the sdOs are usually 
divided into the He-rich and H-rich subclasses with the majority of sdOs 
belonging to the He-rich class. Indeed among the 58 sdOs analysed by 
\citet{stro07}, two-thirds had helium-rich atmospheres (He-sdOs) and \teff\ 
between 40 and 50 kK, while the remaining one-third had hydrogen-rich atmospheres 
(H-sdOs) and were evenly distributed across the temperature range. A similar pattern 
was also found by \citet{nem12}. In addition, the He-sdOs usually have an 
atmosphere that is also enriched in carbon and/or nitrogen. 
The different characteristics of the two spectral classes of sdO are explained by 
their different origins \citep{nap08sdo}. The H-rich sdOs are believed to be 
directly linked to sdBs. They represent the He-shell burning phase, usually 
referred to as the post-EHB phase, directly following the He-core burning stage 
that occurs on the EHB \citep{dor93,han03}. H-sdOs and sdBs are also naturally 
linked by the fact that they both have atmospheres depleted in helium. On the 
other hand, He-sdOs are not believed to be evolutionarily linked to the sdBs, 
but have their own evolutionary channels, such as late flashers or 
white dwarf mergers \citep{dcruz96,bert08,zhang12}.

While many sdB stars have been studied in detail for abundance analyses based on 
UV \citep{bla08,otoole06} and optical spectroscopy \citep{geier13}, very few 
studies have looked at the chemical composition of their hotter sdO 
progeny.  
Because of their relatively cool effective temperatures, sdB spectra can be 
analysed with model atmospheres assuming local thermodynamic equilibrium (LTE), 
which are rather fast and simple to compute. However the LTE approximation no longer holds at the high temperature of sdO stars, and more sophisticated and 
time consuming non-LTE (NLTE) modelling is required. In addition, the hottest 
sdOs show very few metal lines in their optical spectrum, thus scarcely 
revealing information concerning their atmospheric composition. This mostly explains why 
only a handful of spectroscopic analyses of H-rich sdO stars have been 
performed. Up to now three such stars have been studied in detail: AA Dor, Feige 
110, and \bd\ \citep{fle08,klepp11,rauch14,lat13}. 

In this work we add four additional stars to the, as yet, rather small sample of 
thoroughly analysed H-sdOs, namely \feige, \feiget, \agk, and \lsii. For these 
four targets we used high quality optical spectra, UV data from 
\textit{FUSE}\footnote{\textit{Far Ultraviolet Spectroscopic Explorer}} and 
\textit{IUE}\footnote{\textit{International Ultraviolet Explorer}} and  
photometry to perform our comprehensive spectroscopic analysis. 
In Sect.~2 we present these four stars by summarising current literature 
information about the stars. We then describe the optical and UV data used in our 
work in Sect.~3. This is followed by our determination of the atmospheric 
parameters and chemical abundances in Sect.~4 and 5. In Sect.~6 we use 
photometric data to derive spectroscopic distances and characterize the infrared 
excess of \feige. Finally we discuss our results in Sect.~7 and briefly conclude 
in Sect.~8.

\section{The quartet: \feige, \feiget, \agk,\ and \lsii}

Despite the relative brightness of our four stars and the status of optical and 
UV spectrophotometric standard stars for the three stars \feige, \feiget\ and 
\agk\ (\citealt{oke90,turn90}), information regarding their fundamental parameters and 
elemental abundances is rather scarce. In this section we briefly present the 
stars included in our analysis. 

With a $V$ magnitude of 11.14, Feige 34 is the brightest star in our sample. 
\citet{the91} first adopted \teff\ $\approx$ 80~000~K and log $g$ = 5.0 by 
comparing its optical spectrum with that of \bd. The main goal of their study 
was to characterize the infrared excess of the star, which they found to be most 
likely originating from a cooler main-sequence companion. 
The surface gravity of the star was then increased to log $g$ = 
6.8$^{+0.3}_{-0.7}$, based on photometric fitting and the parallax value 
\citep{the95}.
The temperature of the star was revised by \citet{haas97} to 60~000 K using
the ionization equilibrium of iron and nickel lines present in \underline{the}
\textit{IUE} spectrum and line-blanketed non-LTE model atmospheres of the star
\citep{haas96,wer98}. He also quantified the overabundance of these elements to
be 10 and 70 times solar, respectively, and reported the extreme similarity 
between the \textit{IUE} spectra of \feige\ and those of two other hot sdO stars, 
\feiget\ and \lsii. Given the nearly identical UV spectra of the three stars, 
\citet{haas97} concluded that their fundamental parameters and their iron and 
nickel abundances must be essentially the same. 
More recently, \citet{lapal14} reported the detection of X-ray emission from 
Feige 34. This emission could be intrinsic to the star, originating from the companion, 
or possibly both. \citet{val06} observed weak polarization in the H$_{\alpha}$ 
line of \feige, indicating the presence of a weak variable magnetic field ($<$ 
10 kG). However, \citet{land12} reported that most of the magnetic field 
detections (via polarimetry) in hot subdwarfs were later on shown to be spurious 
and no secure detections could be claimed. Additional spectropolarimetric 
observations of \feige\ would be needed to clarify the issue.

Feige 67 was included in the sample of sdO stars analysed by \citet{bauer95}. 
These authors derived \teff\ = 75 000 $\pm$ 5000 K, log $g$ = 5.2 $\pm$ 0.2, and log \nhe 
= $-$1.27 $\pm$ 0.25 based on optical spectroscopy. Using the IUE 
spectrum, they also measured a nitrogen abundance and upper limits for C 
and Si. 
The star was also examined by \citet{beck95a,beck95b,beck95ni} during the 
development of their NLTE nickel and iron model atoms and found these elements 
to be enriched in the star by at least a factor of 10 with respect to the solar 
value. 
\cite{deet00} further analysed ORFEUS~\textsc{ii} and \textit{IUE} spectra of 
the star and confirmed the overabundance of iron and nickel.

\agk\ was first classified as an sdO by \citet{ber78}. This star is a 
spectrophotometric flux standard for\ Hubble Space Telescope (HST) and for the Gaia mission 
\citep{boh01,pan12}. Besides good photometric data, there is little other literature on 
the star itself.  We did not find information concerning its fundamental parameters. 

\lsii\ is our faintest object (V=12.34) and its effective temperature was first 
estimated to be $\approx$ 60 000 K by \citet{schon84} based on the UV 
(\textit{IUE}) continuum slope. Its \textit{IUE} spectrum was further analysed 
by \citet{del92} who did a comparative study of the abundance pattern of \lsii\ 
and the hot subdwarf component of the CPD$-$71$\degr$172 binary system. Although 
their analysis is very detailed, it remained mostly qualitative and no 
abundances were derived. The authors emphasized the similarity of both stars' UV 
spectra. 

Since \feige, \feiget, and \agk\ are among the 46 primary spectrophotometric 
standard candidates for Gaia \citep{pan12}, all three were observed for 
photometric variations and validated against short-term variability to a 
precision level of 4 mmag \citep{marin16}. In addition, \feiget\ and 
\lsii\ were observed to detect short-period pulsations, but nothing was detected 
above 4$\sigma$, corresponding to 0.08\% of the mean brightness \citep{john14}.

\section{Observations}

\subsection{Optical spectroscopy}

We collected a variety of low and medium resolution spectra that are most 
useful to derive atmospheric parameters.  
The four stars have been observed multiple times by one of us (E.M.G.) as part 
of her spectroscopic programmes in recent years \citep{gre08}. A detailed 
description of the three instrumental set-ups can be found in \citet{lat15}, 
hence we only describe the main characteristics of the spectra 
obtained.
For each star we have low ($\Delta\lambda$ = 8.7 \AA) and medium (1.3 \AA) 
resolution spectra taken at the Steward Observatory 2.3 m Bok Telescope on Kitt 
Peak. The low resolution set-up (400/mm grating, 2.5\arcsec\ slit) allows for complete 
coverage of the Balmer series, from 3620 to 6900 \AA, while the medium resolution 
spectra (832/mm grating, 2$^{nd}$, 1.5\arcsec\ slit) cover a bluer 
wavelength range of 3675$-$4520 \AA.  For \feige, we also have spectra taken 
with the MMT blue spectrograph (832/mm grating, 2$^{nd}$, 1.0\arcsec\ slit)   
covering the 4000$-$4950 \AA\ range at a resolution of $\approx$ 1.0 \AA\  with
much higher signal-to-noise ratio (S/N) than the Bok spectra.

 For each set-up described above, the individual spectra were first
cross-correlated against the spectrum with the highest S/N to determine 
initial values for the relative radial velocities (RVs).  Next, the individual 
velocity-corrected spectra were combined to produce a better high-S/N cross-correlation
template, which yielded improved relative RVs.  The process was repeated until
the RVs for all the individual stars converged, resulting in a final, 
optimally combined, high sensitivity spectrum (see also 
\citealt{gre08}). Additional details about the derived radial 
velocities are provided in the Appendix A.

For \feige\ we also have one additional spectrum taken with the blue arm of the 
intermediate dispersion spectrograph and imaging system (ISIS) at the William 
Herschel Telescope\footnote{http://www.ing.iac.es/astronomy/instruments/isis/}. 
The R600B grating combined with a 1$^{\prime\prime}$ slit width provided a 
resolution of 2.0 \AA\ over the 3800$-$5250 \AA\ range.

\agk\ and \lsii\ were observed with the Calar Alto Faint Object 
Spectrograph (CAFOS) on the 2.2 m telescope at Calar Alto. Spectra were obtained 
with the blue$-$100 and red$-$100 grisms and the slit width of 100 $\mu$m 
provided a resolution of 5.0 \AA. The blue and red spectra cover the 3600$-$9000 
\AA\ range. 

All the spectra were reduced and extracted using standard IRAF routines for 
long-slit spectroscopy.

\subsection{UV spectroscopy}

\begin{table}[b]
\caption{Ultraviolet data retrieved from the MAST archive}\label{uvdata}
\centering
\scriptsize
\begin{tabular}{llccc}
\hline
\hline
Telescope & Star & Dataset & Exp. Time (s) & Obs. date \\
\hline 

    \multirow{7}{*}{\textit{IUE}}& 
                               Feige 34 & SWP17349 & 7200 & 03/07/1982 \\
                               & & SWP20125 & 4200 & 02/06/1983 \\
     & Feige 67 & SWP20488 & 10800 & 20/07/1982 \\ 
     & \agk & SWP17344 & 9000 & 03/07/1982 \\  
               & \lsii\ & SWP18069 & 13800 & 23/09/1982 \\
                     & & SWP51580 & 15000 & 19/08/1994 \\   
                               & & SWP52064 & 14520 & 08/09/1994 \\              
                                
  \hline   
  
    \multirow{8}{*}{\textit{FUSE}}& 
                                Feige 34 & M1052201000 & 5590 & 21/03/2003 \\
                               &  & M1080302000 & 1847 & 03/05/2000 \\
                               &  & P2040701000 & 6307 & 10/02/2001 \\ 
        & Feige 67 &  M1080701000 & 5140 & 29/04/2000 \\ 
                           & \agk\ &    M1080501000  &  6864 & 27/11/2000 \\
                               &  & M1052103000  & 5087 & 21/01/2003 \\
                               &  & M1052104000 & 6029 & 08/02/2003 \\     
         & \lsii\ &  P2051501000         & 3525 & 29/05/2004 \\                  
              
\hline
\end{tabular} \\
\end{table}

All of our stars have been observed with the short-wavelength prime (SWP) camera 
in the high-dispersion mode on board the 
\textit{IUE} satellite. The spectra cover a wavelength range of 1150 to 2000 
\AA\ with a resolution of $\approx$ 0.15 \AA. 
The spectra retrieved from the Barbara A. Mikulski Archive for Space Telescopes 
(MAST) are listed in Table \ref{uvdata}. When multiple observations were 
available for one star, we combined the spectra and merged the orders using the 
routines provided in the IUEDAC software libraries. 

The stars were also observed by the \textit{FUSE} satellite, which offers a 
complementary wavelength range (905$-$1187 \AA) to the \textit{IUE} spectra and a higher resolution ($\approx$ 0.06 \AA). We retrieved the observations 
available from the archive (see Table \ref{uvdata}) that were taken through the 
LWRS (30 $\times$ 30$^{\prime\prime}$) aperture and recorded in histogram mode. 
The only exception is \lsii,\ which has only a MDRS (4 $\times$ 
20$^{\prime\prime}$) observation. Each observation consists of multiple 
exposures. We used the fully calibrated and extracted spectra provided in the 
archive for each exposure and detector segment. The different exposures were 
cross-correlated and co-added for the eight segments. This way we obtained a 
combined spectrum for each segment. At this stage we did not include individual 
exposures if they had a discrepant flux level; this could happen, for example, 
if the target was not well centred in the aperture. The LiF1B spectra show a 
depression of flux between 1130 \AA\ and 1170 \AA,\ that is more or less 
important depending on the observation. For this reason, we only kept the region 
longward of 1170 \AA\ for this segment. 
Given that there is significant overlap in the wavelength covered by the 
different segments and that their wavelength calibration can slightly differ, we 
considered the segments with the best S/N over a given wavelength range to obtain our final \textit{FUSE} spectra. This way we avoided smearing 
that would be produced by co-adding more than one segment over a wide wavelength 
interval. For example, the final co-added \textit{FUSE} spectrum of \feige\ was 
obtained by considering and merging the following spectral regions: SiC1B 
(910$-$990 \AA), LiF1A (990$-$1080 \AA), SiC1A (1080$-$1089 \AA), LiF2A 
(1089$-$1180 \AA), and LiF1B (1175$-$1187 \AA).

\section{Atmospheric parameters}
We estimated the fundamental parameters (\teff\ and log $g$) and helium abundance by simultaneously fitting the optical H and He lines. We used a 
$\chi^2$ minimization procedure that relies on the method of Levenberg-Marquardt 
\citep{ber92}. Normalized lines of both the observed and model spectra are thus 
compared. This is a standard method used to estimate fundamental parameters of 
hot subdwarfs. 
In the case of hot sdOs, such as those analysed here, both NLTE 
effects and line blanketing by heavy elements have to be taken into account to derive reliable temperatures and resolve the Balmer line problem 
\citep{ber93,wer96}. 
\citet{lat15} showed that stellar atmosphere models with enhanced individual 
metal abundances (10 times solar) are required to solve the Balmer line problem 
observed in the hot sdO star \bd. The increased abundances compensate for 
missing opacities in the models.

For this work we used similar NLTE line-blanketed model atmospheres computed 
with the public codes TLUSTY and SYNSPEC \citep{lanz95,lanz03}. We computed a 
model grid covering the following parameter range: 50~000 K $<$ \teff\ $<$ 
68~000 K by 2000 K steps, 4.8 $<$ log $g$ $<$ 6.4 by 0.2 dex steps, and $-$4.0 
$<$ log \nhe\ $<$ 0.0 by 0.5 dex steps. The models include line blanketing from 
the following elements: H, He, C, N, O, Mg, Si, S, Fe, and Ni.

\begin{figure*}[t]
\includegraphics[scale=0.37,angle=270]{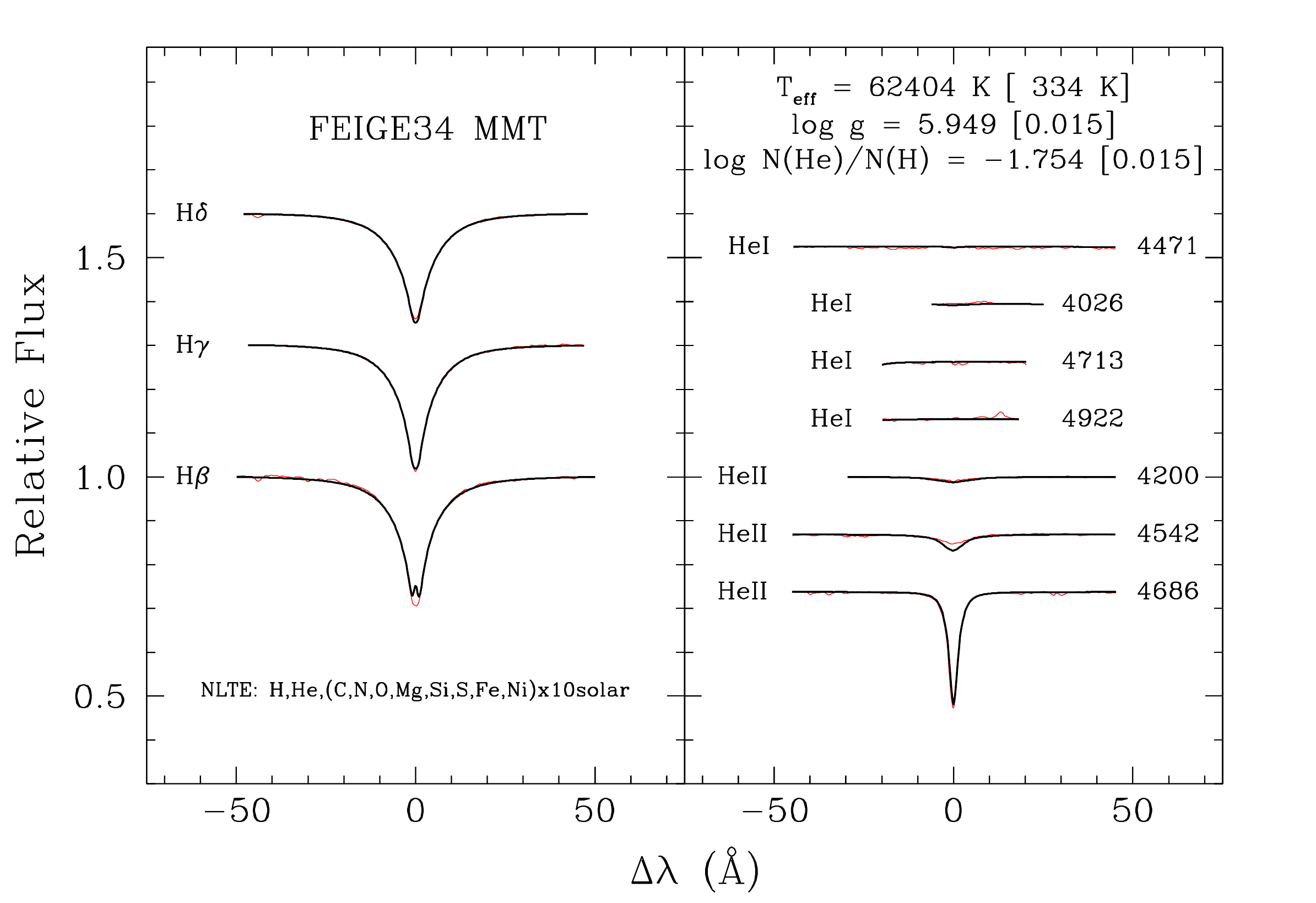}
\includegraphics[scale=0.37,angle=270]{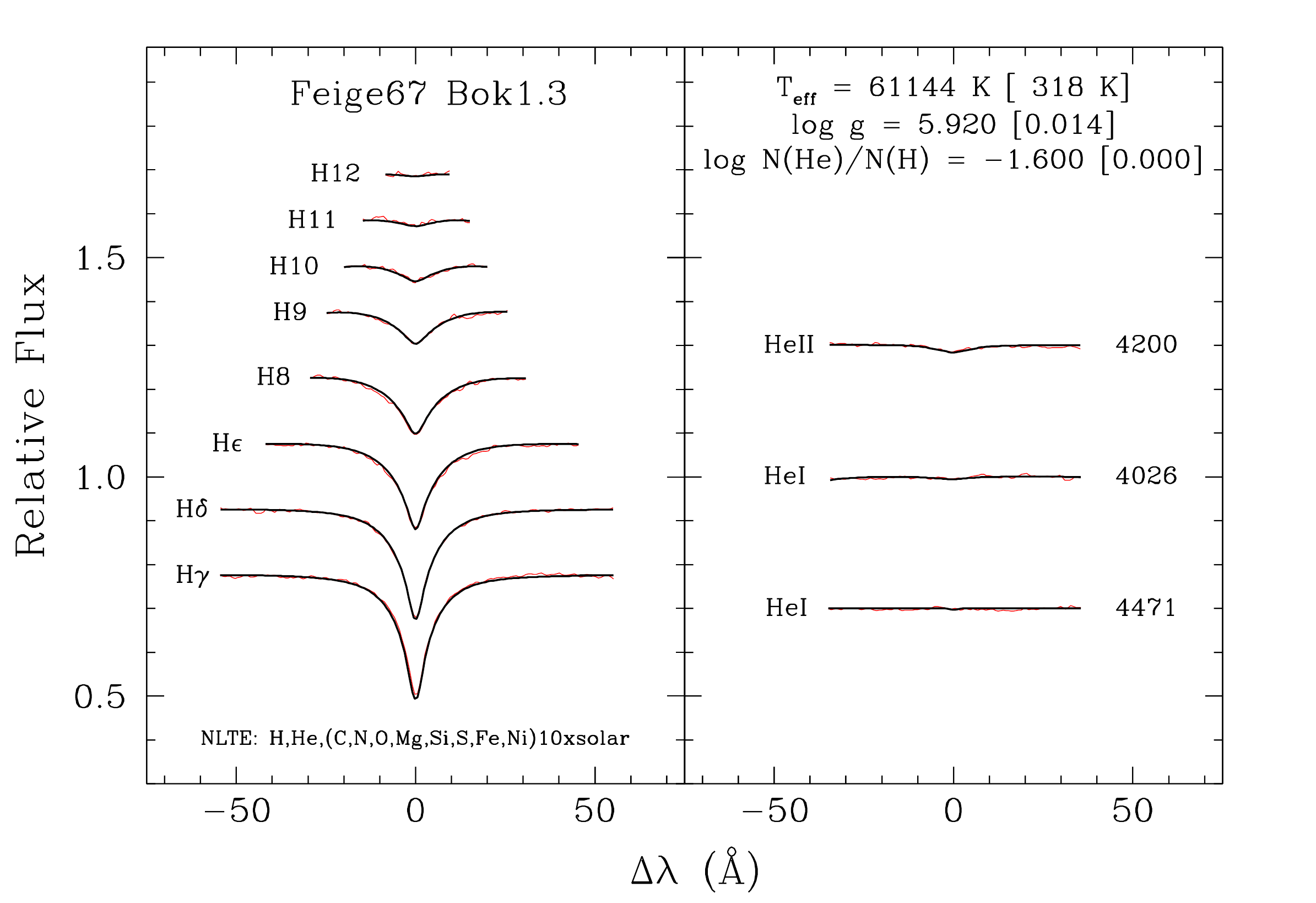}
\includegraphics[scale=0.37,angle=270]{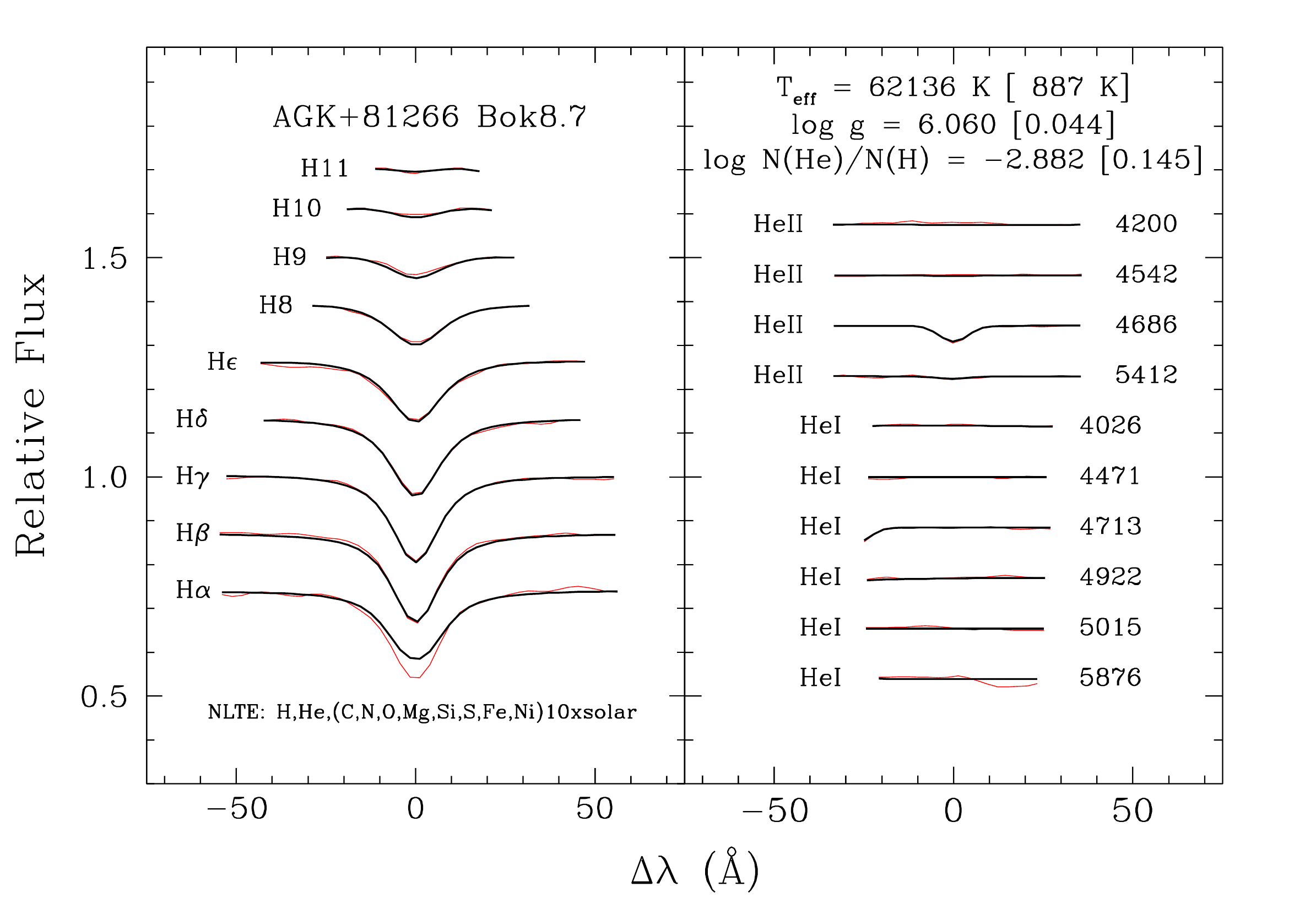}
\includegraphics[scale=0.37,angle=270]{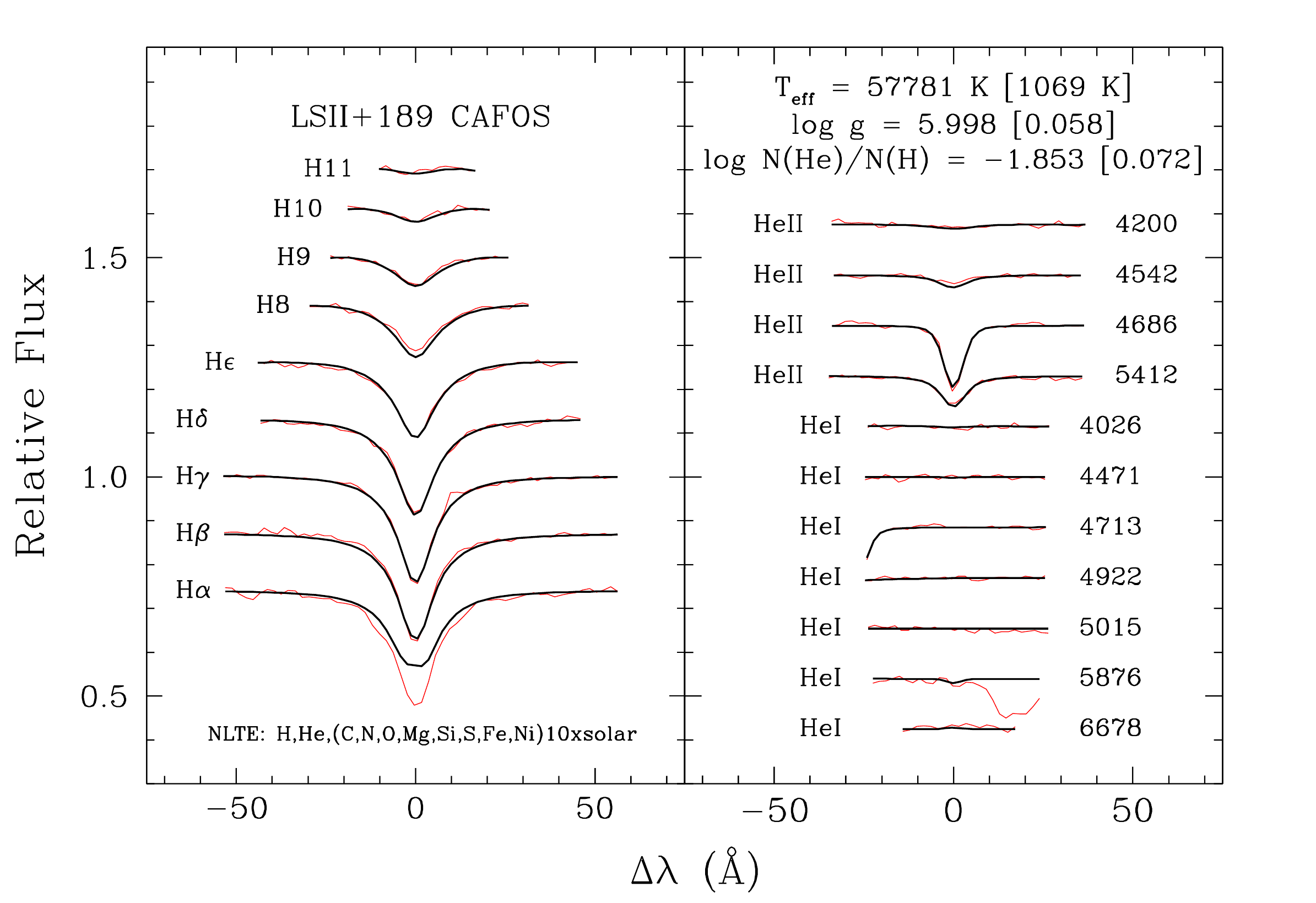}
\caption{Best fits to some of our optical spectra. The observed spectral lines 
are 
  shown in red, while the modelled lines are in black.}
\label{fit}
\end{figure*}

\begin{table*}
\center
\caption{Fundamental parameters derived from optical spectroscopy}
\label{param}
\small
\begin{tabular}{llccc}
\hline
\hline
Star & Spectrum & \teff\ (K) & log $g$ & log \nhe \\
\hline 

    Feige 34 & MMT 1.0\AA & 62 400 $\pm$ 330 & 5.949 $\pm$ 0.015  & $-$1.754 
$\pm$ 0.015 \\     
     & BOK 1.3\AA & 62 317 $\pm$ 300 & 6.032 $\pm$ 0.012 & 
$-$1.8\tablefootmark{a}  \\
     & ISIS 2.0\AA & 61 974 $\pm$ 304 & 5.980 $\pm$ 0.013  & $-$1.765 $\pm$ 
0.021 \\
     & BOK 8.7\AA & 63 510 $\pm$ 820 & 5.984 $\pm$ 0.035  & $-$1.842 $\pm$ 0.039 
\\
     Feige 67 & BOK 1.3\AA & 61 114 $\pm$ 318 & 5.920 $\pm$ 0.014 & 
$-$1.6\tablefootmark{a}  \\
              & BOK 8.7\AA & 61 000 $\pm$ 723 & 5.981 $\pm$ 0.041 & $-$1.594 
$\pm$ 0.038 \\ 
   \agk\ & BOK 1.3\AA & 62 590 $\pm$ 538 & 6.067 $\pm$ 0.021 & 
$-$2.9\tablefootmark{a}  \\
    &  CAFOS 5.5\AA & 60 859 $\pm$ 1182 & 6.093 $\pm$ 0.063 & $-$2.986 $\pm$ 
0.200  \\
    &  BOK 8.7\AA & 62 136 $\pm$ 900 & 6.060 $\pm$ 0.044  & $-$2.882 $\pm$ 0.140 
\\
     
   \lsii & BOK 1.3\AA &  61 390 $\pm$ 350 & 5.992 $\pm$ 0.015 & 
$-$1.9\tablefootmark{a} \\
                 &  CAFOS 5.5\AA & 57 767 $\pm$ 1067 & 5.996 $\pm$ 0.058 & 
$-$1.851 $\pm$ 0.071 \\
                 & BOK 8.7\AA & 60 884 $\pm$ 996 & 6.024 $\pm$ 0.056  & $-$1.890 
$\pm$ 0.056 \\   
                                                
  \hline  
\end{tabular}
\tablefoot{
\tablefoottext{a}{Because of the absence of visible He lines in the Bok 1.3\AA\ 
spectral range, the abundance was kept fixed.}
}
\end{table*}

Fig.~\ref{fit} shows examples of our best fits to the Balmer and helium lines 
and Table \ref{param} summarizes our results. The uncertainties are statistical 
and mostly reflect the S/N of the spectra.
For instance, our CAFOS lower S/N spectra fits resulted in higher 
uncertainties. 
An additional source of uncertainties can arise from various resolutions and 
wavelength coverage of the fitted spectra. However, the parameters we determined 
from the different spectra are remarkably consistent. The four spectra of 
\feige\ give temperature values consistent within 1600 K.

Our results confirm Haas' \citeyear{haas97} analysis that \feige, \feiget, and 
\lsii\ have very similar atmospheric parameters. The helium abundances are also 
similar and vary by less than 0.3 dex.
To this group of stars we now add \agk,\ which shares equally comparable \teff\ 
and log $g$. As yet, the star only differs by its helium abundance, which is about 10 
times lower.

The optical spectral lines are well reproduced by our best fitting models with 
the exception of the cores of the lowest Balmer lines, i.e. H$\alpha$ and 
H$\beta$. 
At such high temperatures the cores of these lines turn into emission, but the 
low resolution of our spectra smears out these features.
To inspect the observed line profiles we retrieved some HIRES spectra of \feige\ 
from the Keck Observatory 
Archive\footnote{http://www2.keck.hawaii.edu/koa/public/koa.php}. 
A comparison between our models and the high resolution observations showed that 
the models predict NLTE emission cores to be stronger than the observed cores (see 
Fig.~\ref{halpha}). At lower resolution, the smearing out of the emission 
features leads to shallower line profiles, thus explaining the discrepancies 
observed in our fits for H$\alpha$ and H$\beta$. 

The parameters we derived for our stars are in good agreement with the values 
stated in the literature. The temperatures match well that derived by 
\citet{haas97} and our surface gravities are between the value (log $g$ = 5.2) 
determined by \citet{bauer95} for \feiget\ and the higher value (log $g$ = 6.8) 
estimated by \citet{the95} for \feige.

\section{Abundance analysis}

\subsection{Rotational broadening}

\begin{figure}
\begin{center}
\resizebox{\hsize}{!}{\includegraphics{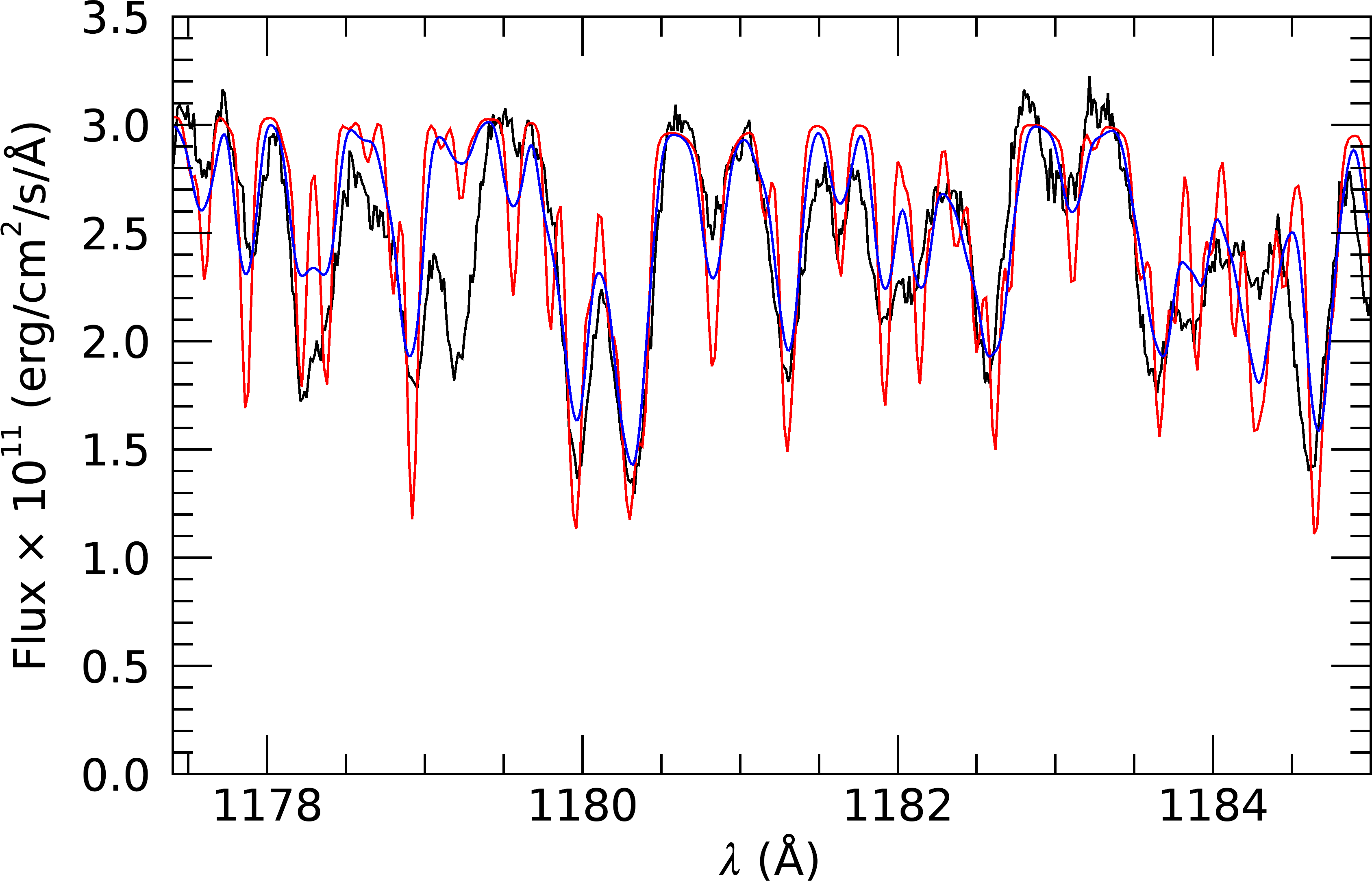}}
\caption{ Comparison of \textit{FUSE} LiF1B spectrum of Feige 34 with model spectra 
without
rotational broadening (red) and with \vsini\ = 25 km s$^{-1}$ (blue). }
\label{vsini}
\end{center}
\end{figure}

A comparison of the \feige\ \textit{FUSE} spectrum with model spectra readily 
provided indications of line broadening; the modelled lines had systematically 
sharper cores than the observed lines. Because the metal lines in the 
\textit{FUSE} spectrum are heavily blended, it is not straightforward to use 
well-defined and isolated lines for measuring the rotational broadening. Instead 
we selected a wavelength range ($\approx$1165$-$1185 \AA) for which most of the 
observed lines were present in our models. This range is covered by two of the 
\textit{FUSE} segments (LiF2A and LiF2B) and the \textit{IUE} spectrum 
and is mostly dominated by Ni~\textsc{V}$-$\textsc{VI} lines. We performed a fit 
of this region using a model grid in which the nickel abundance and rotation 
velocity were varied\footnote{The synthetic spectra are convolved 
with a Gaussian to match the \textit{FUSE} (0.06 \AA) and \textit{IUE} (0.15\AA) 
instrumental resolution.}, and we obtained \vsini\ values between 25 and 30 km 
s$^{-1}$ for the different spectra.  We repeated the fitting exercise 
for the three other stars, which also show signs of additional broadening.
Consistent with the remarkably similar nature of the \textit{FUSE} spectra of 
our stars, we obtained values of \vsini\ between 20 and 35 km s$^{-1}$ for all 
four stars and their different spectra (\textit{FUSE} and \textit{IUE}).
Figure \ref{vsini} shows part of the fitted range for the LiF1B segment of 
\feige\ along with synthetic spectra that have \vsini\ = 25 km s$^{-1}$ and no 
rotational broadening.

Rotational broadening has not been reported in most of the previous work on 
these stars. 
However the work of \citet{del92} reported the presence of line broadening in 
\lsii\ and CPD$-$71$\degr$172B. These authors were able to distinguish the effect \underline{with} the 
Fourier filtering technique they applied to the \textit{IUE} data to optimally 
remove the noise. They concluded that it could be explained either by stellar 
rotation or a high microturbulence. \citet{beck95a} also mentioned the presence 
of rotational broadening in the \textit{IUE} spectrum of \feiget. 

%

To explore the possibility of microturbulence, we fit four nickel lines that are observed in the FUSE and IUE spectra of Feige 34. These lines-- the Ni VI lines at 1179.920,
1179.966, 1180.300, and 1180.388 \AA--are not blended with any other metal lines. Although the transitions start from energy levels above 300,000 cm$^{-1}$, they have strong oscillator strengths that range from log gf = 0.033 to 0.467. The best fit between the observed data and model was determined by considering the microturbulent velocity ($v_{\rm micro}$) and the nickel abundance as free parameters. The microturbulent velocity is only considered in the calculations of the synthetic spectra (SYNSPEC), while the nickel abundance is considered explicitly in the calculations of the stellar atmosphere models (TLUSTY). A grid of synthetic spectra was computed by considering microturbulent velocities in the range of 0 $\leq$ $v_{\rm micro}$ $\leq$ 22 km s$^{-1}$, in steps of 2 km s$^{-1}$, and nickel abundances in the range of $-4.9 \leq \log N(Ni)/N(H) \leq -3.4$, in steps of 0.3 dex. The microturbulent velocity is assumed constant throughout the atmosphere. We estimated that if we considered the microturbulence as an additional broadening in SYNSPEC only, we could overestimate the nickel abundance by 0.2 dex and underestimate the microturbulent velocity by 1.1 km s$^{-1}$.

The best fit of the FUSE and IUE data gives $v_{\rm micro}$ = 18 $\pm$ 2 km s$^{-1}$ and log $N$(Ni)/$N$(H) = $-$4.7 $\pm$ 0.1. The uncertainties are given by the standard deviation of the measurements. The lower nickel abundance is explained by the fact that the nickel lines are saturated and because the effect of the microturbulence widens the wavelength range covered by the absorption lines and reduces the saturation. A lower nickel abundance is then needed to reproduce the lines. Given that the microturbulent velocity that we measured for Feige 34 is very high and implausible for such a compact star, we favour the rotational broadening option. For our abundance analysis, we adopted a value of \vsini\ = 25 km s$^{-1}$ for all the stars and have not included any microturbulent velocity.

\subsection{Chemical composition}

\begin{figure*}
\begin{center}
\resizebox{\hsize}{!}{\includegraphics[angle=90]{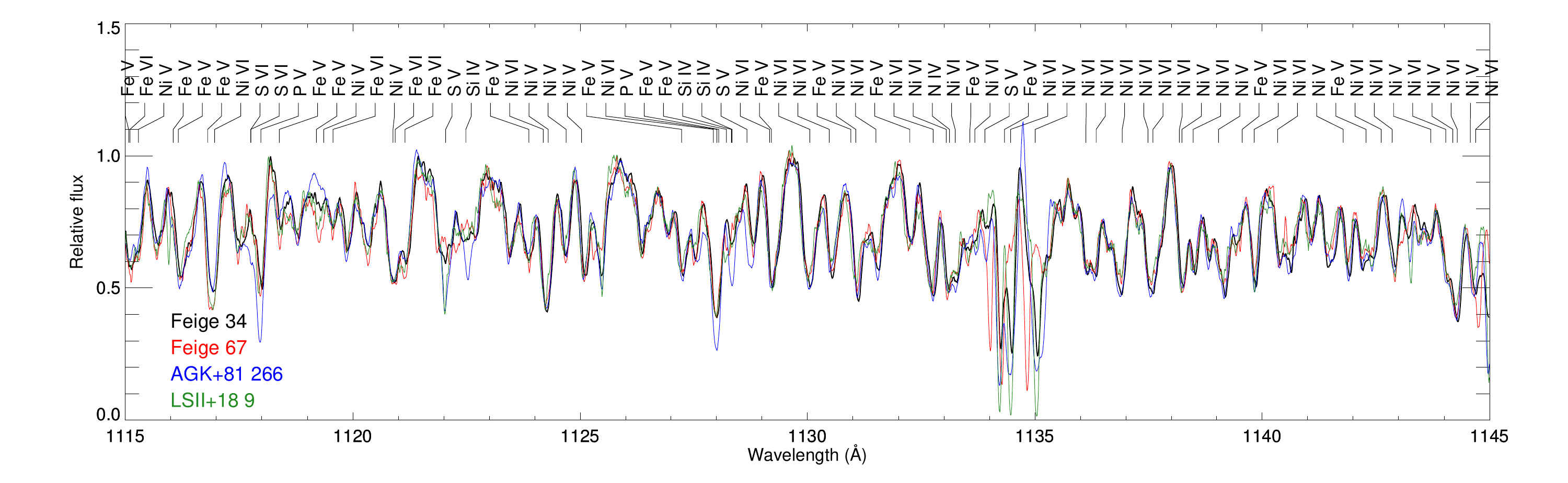}}
\resizebox{\hsize}{!}{\includegraphics[angle=90]{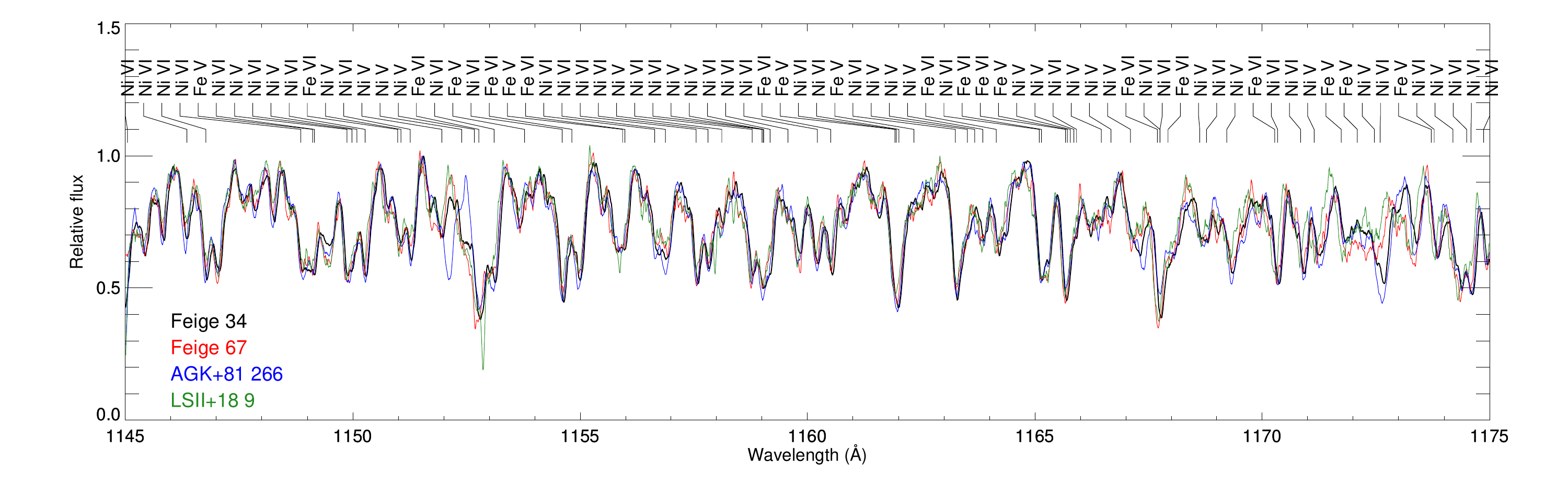}}
\caption{Comparison between the \textit{FUSE} spectra of our four stars. The 
observed spectra are smoothed with a moving average over three 
pixels for better visualization and are shifted to the rest wavelength. 
Also shown are labels identifying the strongest lines in our linelist. }
\label{fusecomp}
\end{center}
\end{figure*}

\begin{figure*}
\begin{center}
\resizebox{\hsize}{!}{\includegraphics[angle=90]{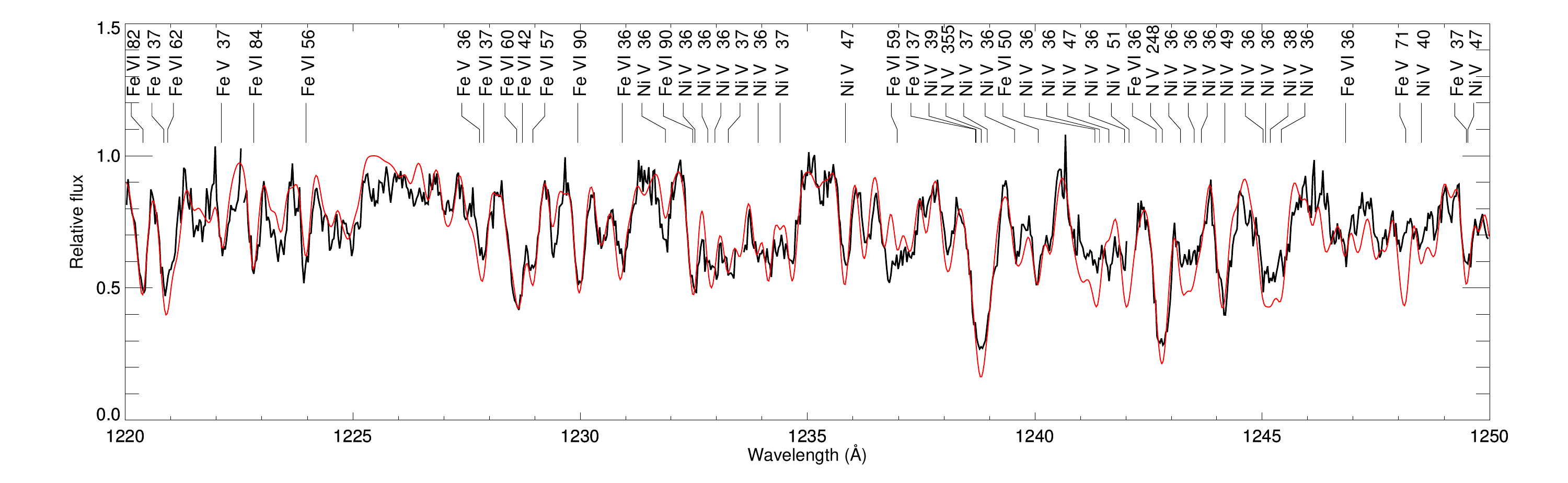}}
\resizebox{\hsize}{!}{\includegraphics[angle=90]{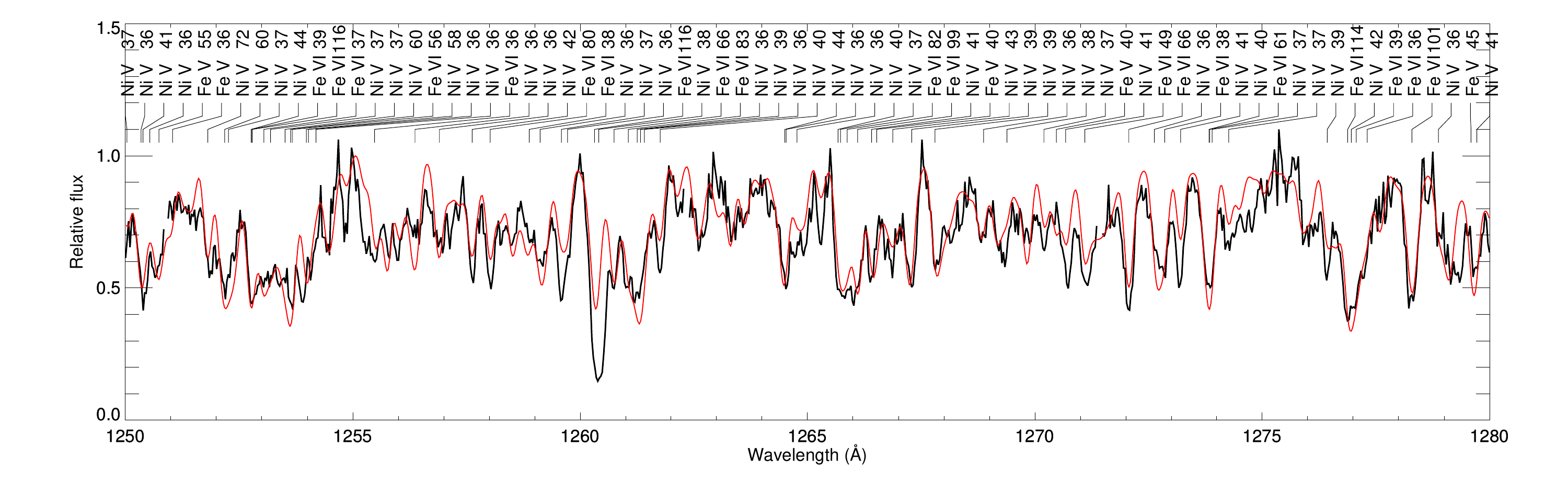}}
\caption{Our model spectrum (red) compared to the \textit{IUE} spectrum of 
Feige 34 in the range used for fitting the Ni lines and \ion{N}{v} doublet. Lines 
with equivalent width larger than 36m\AA\ (given in the labels in m\AA) are 
indicated. The 1259.5 and 1260.5 \AA features are due to \ion{S}{ii} and \ion{Si}{ii} interstellar lines.}

\label{iueni}
\end{center}
\end{figure*}

The UV spectral range is dominated by many strong iron and nickel (mostly 
\textsc{v} and \textsc{vi}) lines. As outlined by \citet{haas97}, the 
\textit{IUE} spectra of \feige, \feiget, and \lsii\ are extremely similar, as 
is the spectrum of \agk, which shows the same strong features of iron and 
nickel. The resemblance is even more striking when comparing the \textit{FUSE} 
spectra of our stars with their better resolution and S/N, many features 
clearly appear to be identical in all the stars. 

While absorption from iron and nickel dominates the spectra of our stars, the 
absorption lines from lighter elements such as carbon, oxygen, and nitrogen are 
much more discreet, suggesting that these lighter elements are depleted. It is 
with respect to these lighter elements that \agk\ stands out from the other 
stars, showing a somewhat different pattern. For example, it has weaker lines of 
oxygen and nitrogen, while silicon, phosphorus, and sulfur show stronger 
features. 
This is illustrated in Fig.~\ref{fusecomp}, where we overplotted the 
\textit{FUSE} spectra of our four stars over the 1115$-$1175 \AA\ range. The 
spectral similarity between the stars is very obvious as are the 
differences in the \ion{P}{v} ($\lambda\lambda$ 1118, 1128), S ($\lambda\lambda$ 
1117.7 1128.7,1133.9), and \ion{Si}{iv} $\lambda$1128.3 lines in \agk. The three 
strong absorption features around 1135 \AA\ are \ion{N}{i} interstellar (IS) 
lines, and the spectra of \agk\ and \lsii\ additionally show prominent 
\ion{Fe}{ii} IS absorption at 1122 and 1145 \AA.
The \textit{FUSE} range at shorter wavelength contains a fair amount of 
interstellar transitions, originating mostly from molecular H$_2$, \ion{H}{i}, 
and neutral and singly ionized atoms such as C, N, O, Ar, and Si. Our stars 
are affected differently by interstellar contamination, H$_2$ seems to be almost 
absent from the line of sight of \feiget, while it shows strong features in 
\lsii, which also has stronger IS \ion{H}{I} absorption. Similarly,
the interstellar \ion{Ca}{i} $\lambda$3933 line in the Bok spectra
is undetectable in \feiget, extremely weak in \feige, and rather strong in 
\agk\ and \lsii.

For the abundance analysis we used a common set of parameters determined from 
the optical analysis (62 000 K, log $g$ = 6.0) to build our model grids. The 
helium abundances were fixed to log \nhe\ = $-$1.8 for \feige, \feiget, and \lsii, 
and log \nhe\ =$-$2.9 was used for \agk. We built grids where one metallic element 
was varied at a time. The models computed included C, N, O, Si, S, Fe, and Ni, 
with abundances previously derived or estimated from preliminary analysis. 
Since iron and nickel are the main contributors to the opacity, both in terms of 
emergent spectra and atmospheric structure, their abundances were evaluated 
first, followed by the lighter metals.
The resulting abundances, radial velocities, and averaged atmospheric 
parameters for the four stars are presented in Table \ref{abund}. The individual 
elements are discussed below.

\begin{table*}
\caption{Atmospheric parameters and chemical composition of our sample of hot 
sdOs.}
\label{abund}
\centering
\small
\begin{tabular}{lccccccc}
\hline
\hline
   & \feige & \feiget  & \agk & \lsii & \citet{haas97}\tablefootmark{a} & 
\citet{bauer95}\tablefootmark{b} &  \\
\hline
\teff\ \tablefootmark{c} (K) & 62 550 $\pm$ 600         & 61 050 $\pm$ 520  &  
61 860 $\pm$ 870  & 60 010 $\pm$ 800 & 60 000 $\pm$ 4000 & 75 000 $\pm$ 5000    
& \\
log $g$\tablefootmark{c} (cm s$^{-2}$) & 5.99 $\pm$ 0.03 & 5.95 $\pm$ 0.03  & 
6.07 $\pm$ 0.04  & 6.00 $\pm$ 0.06  & ... & 5.2 $\pm$ 0.2 & \\
RV\tablefootmark{d} (km s$^{-1}$) &  11.0 $\pm$ 7.7  &  27.0 $\pm$ 7.8  &  
$-$27.8 $\pm$ 7.8 &  $-$49.5 $\pm$ 12.3  & ... & ... & \\
\hline
 Element & \multicolumn{6}{c}{Abundance (log $N$(X)/$N$(H))}  &  
Solar\tablefootmark{e}\\
\hline        
 He\tablefootmark{d}  & $-$1.79 $\pm$ 0.04 &  $-$1.59 $\pm$ 0.04 & $-$2.9 $\pm$ 
0.2  & $-$1.87 $\pm$ 0.07  & ... & $-$1.27 $\pm$ 0.25 &  $-$1.07\\
 C   & $\la$ $-$6.7 & $\la$ $-$8.0 & $\la$ $-$6.7  &  $\la$ $-$6.7 & ... & $\la$ 
$-$4.50 &       $-$3.56\\
 N   & $-$4.9 $\pm$ 0.3 & $-$4.9 $\pm$ 0.3  & $-$6.0 $\pm$ 0.3  & $-$4.9 $\pm$ 
0.3 & ... & $-$4.85 $\pm$ 0.5 & $-$4.17 \\
 O   & $-$5.5 $\pm$ 0.5  & $-$5.5 $\pm$ 0.5  & $\la$ $-$6.4  & $-$5.5 $\pm$ 0.5 
& ... & ... & $-$3.31\\
 Mg  & $\la$ $-$5.0   &  $\la$ $-$5.0  & $\la$ $-$5.0  & $\la$ $-$5.0 & ... & 
... & $-$4.40\\
 Si & $-$6.2 $\pm$ 0.3  &  $\la$ $-$6.2  &  $-$5.0 $\pm$ 0.3  &   $-$6.2 $\pm$ 
0.3 & ... & $\la$ $-$5.5 & $-$4.49\\
 P &  $-$6.7 $\pm$ 0.5 & $-$6.7 $\pm$ 0.5  &  $-$5.5 $\pm$ 0.5 & $-$6.7 $\pm$ 
0.5  & ... & ...&  $-$6.59\\ 
 S & $-$5.6 $\pm$ 0.3   & $-$5.6 $\pm$ 0.3  & $-$5.3 $\pm$ 0.3  & $-$5.6 $\pm$ 
0.3 & ... & ... & $-$4.86\\
 Cr & $\la$ $-$5.3 & $\la$ $-$5.3  & $\la$ $-$5.3  &  $\la$ $-$5.3 & ... & ...&  
$-$6.36 \\
 Mn & $\la$ $-$5.6 & $\la$ $-$5.6  & $\la$ $-$5.3  & $\la$ $-$5.6  & ... & ... & 
$-$6.57 \\
 Fe  & $-$3.1 $\pm$ 0.3   & $-$3.1 $\pm$ 0.3  & $-$3.1 $\pm$ 0.3  &  $-$3.1 
$\pm$ 0.3 & $-$3.5 & ... & $-$4.50 \\ 
 Co & $\la$ $-$5.8 & $\la$ $-$5.8  &  $\la$ $-$5.8  &  $\la$ $-$5.8 & ... & ... 
& $-$7.01 \\
 Ni  &   $-$4.0 $\pm$ 0.3   & $-$4.0 $\pm$ 0.3  &  $-$4.0 $\pm$ 0.3 & $-$4.0 
$\pm$ 0.3 & $-$3.9 & ... & $-$5.77 \\       
\hline
\end{tabular}\\
\tablefoot{
\tablefoottext{a}{For \feige, \feiget, and \lsii, parameters such as log $g$ and 
abundances of He, C, and N were kept fixed to the values derived by 
\citet{bauer95} for \feiget.}
\tablefoottext{b}{For \feiget.}
\tablefoottext{c}{Our results are the averaged values of the different 
fits presented in Table \ref{param}. }\
\tablefoottext{d}{Derived from the optical spectra.}\
\tablefoottext{e}{\cite{asp09}.}
}
\end{table*}

\paragraph{Carbon}
The only visible carbon lines are the resonance \ion{C}{iv} doublet in the 
\textit{IUE} spectrum. However these transitions are also common in the 
interstellar medium and since none of our stars have a high radial velocity, we 
do not resolve possible blending of photospheric and IS components. 
In Feige 34 and Feige 67 the lines are shifted with respect to the other stellar 
features, while they are found at a position consistent with a stellar origin 
for \agk\ and \lsii. 
Therefore we only estimated upper limits for the carbon abundance. These upper limits do not produce carbon features in the \textit{FUSE} range 
and no features are seen in the observed spectra either.

\paragraph{Nitrogen}
The \textit{IUE} spectrum features the \ion{N}{v} doublet 
($\lambda\lambda$1238.8, 1242.8) and the \ion{N}{iv} ($\lambda$1718) line while 
four components of the \ion{N}{iv} multiplet around 922$-$924 \AA\ and 
\ion{N}{iv} $\lambda$955 can be distinguished in the \textit{FUSE} spectrum. The 
\ion{N}{iii} $\lambda$991 seems to be present only in \feiget.  None of our stars display P Cygni profile characteristics in 
their \ion{N}{v} doublet, which would indicate the presence of a stellar wind. 
A comparison between our model spectrum and the \textit{IUE} spectrum of \feige\ 
around the \ion{N}{v} doublet range can be seen in Fig.~\ref{iueni}. The 
strongest lines are indicated at the top, along with their estimated equivalent 
width as approximated by SYNSPEC. Since most of the absorption features are 
blends, the equivalent width of the individual lines gives an indication of 
their relative strength.

\begin{figure*}
\begin{center}
\resizebox{\hsize}{!}{\includegraphics[angle=90]{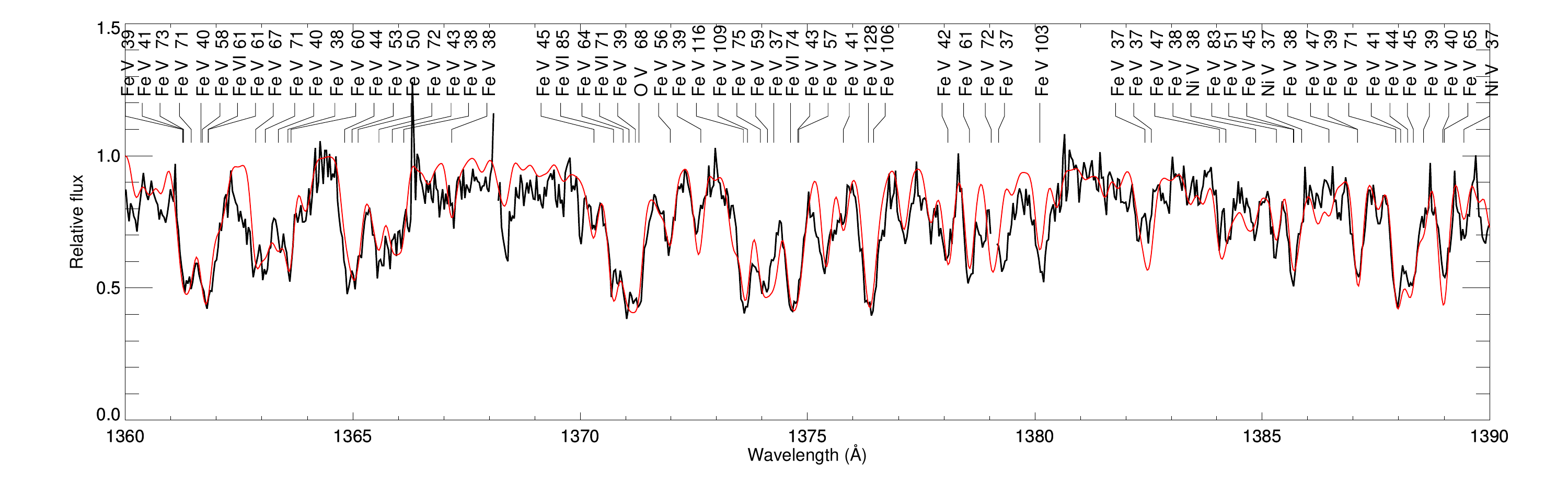}}
\resizebox{\hsize}{!}{\includegraphics[angle=90]{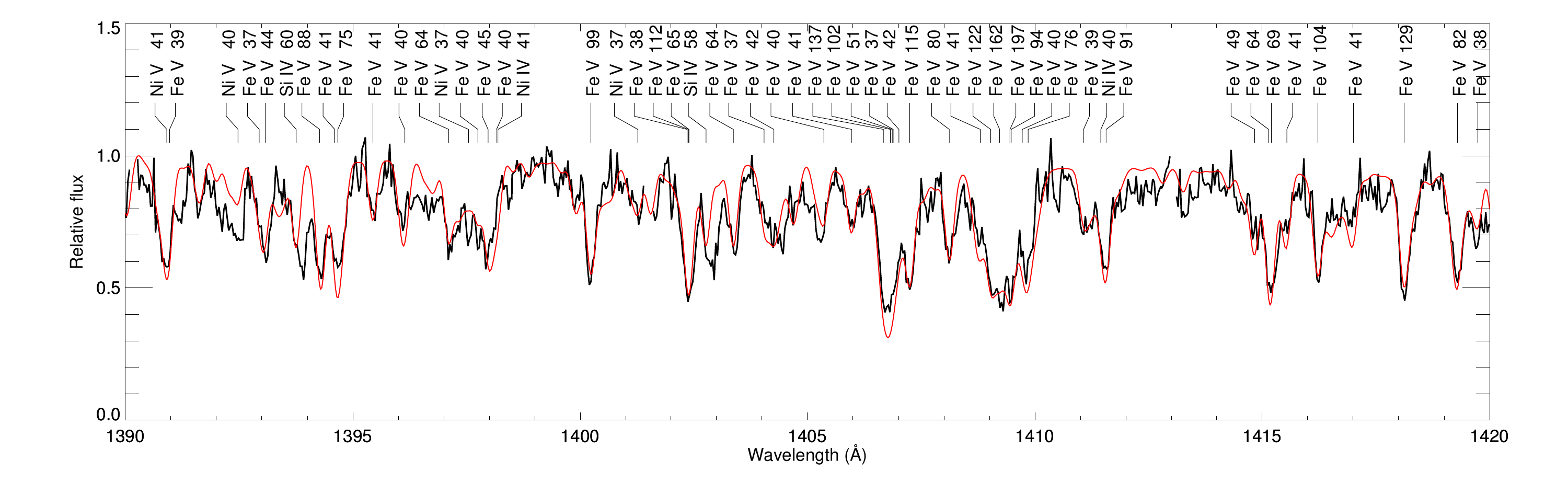}}
\resizebox{\hsize}{!}{\includegraphics[angle=90]{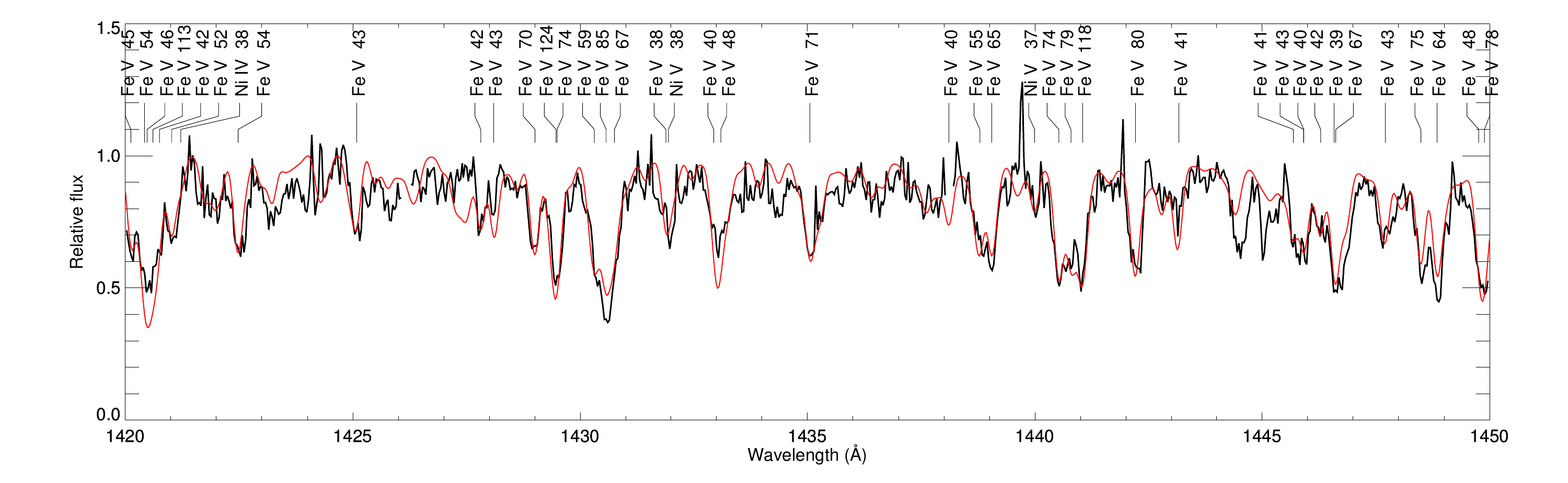}}
\resizebox{\hsize}{!}{\includegraphics[angle=90]{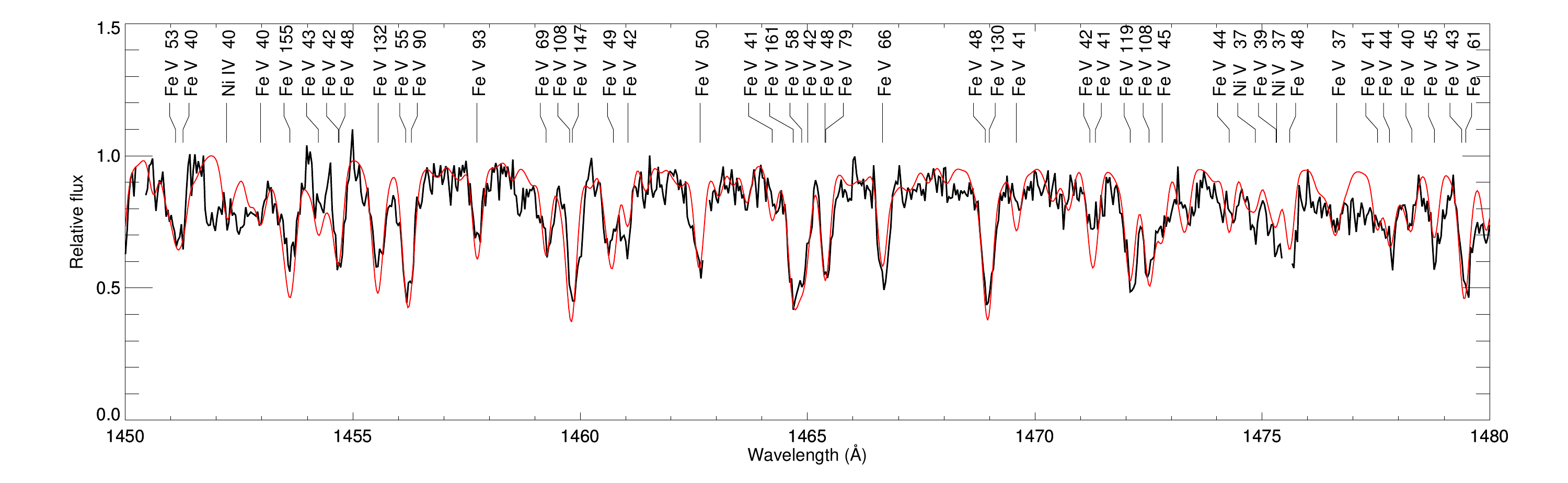}}
\caption{ Same as Fig.~\ref{iueni} but showing the range used for determining 
the iron abundance. The \ion{O}{v} $\lambda$1371 line and \ion{Si}{iv} doublet 
($\lambda$1394, 1403 \AA) are also found in this range. }
\label{iuefe}
\end{center}
\end{figure*}

\paragraph{Oxygen}
The oxygen features are relatively weak given the low abundance of this element 
and most of these features are blended with Fe and Ni lines. The \ion{O}{v} 1371\AA\ line 
is severely blended with two especially strong \ion{Fe}{vi} lines (see 
Fig.~\ref{iuefe}) while the \ion{O}{iv} lines ($\lambda\lambda$ 1338, 1342, 
1343) in the \textit{IUE} range are not particularly strong in our best 
\textit{IUE} spectrum (\feige). The \ion{O}{vi} doublet ($\lambda\lambda$1031.9, 
1037.6) is present in the \textit{FUSE} range, but is blended with the 
\ion{Fe}{v} $\lambda$1031.9 and $\lambda$1032.0 lines and the \ion{Ni}{vi} 
$\lambda$1037.6 line.
The oxygen features in \agk\ are even weaker and suggest that the \ion{Fe}{vi} 
lines in the 1370 \AA\ complex are likely too strong in our models, as the 
absorption remains too strong even after the contribution from \ion{O}{v} 
$\lambda$1371 disappears. 
For this reason our abundances of oxygen have a higher error, and we stated only 
an upper limit in the case of \agk.

\paragraph{Magnesium} 
The absence of Mg \textsc{iii} and \textsc{iv} features in the \textit{IUE} 
spectrum suggests an upper limit of log $N$(Mg)/$N$(H) $\la$ $-$5.0. 

\paragraph{Silicon}
All the silicon features are relatively weak. In \feige, the \ion{Si}{iv} 
doublet ($\lambda$1394, 1403 \AA) appears to be redshifted with respect to the 
photospheric lines (see Fig.~\ref{iuefe}). Like the \ion{C}{iv} doublet, this feature 
appears to be dominated by an IS component. The \ion{Si}{iv} $\lambda$1128 is 
visible in \agk, but rather weak in the other stars.

\paragraph{Phosphorus}
The resonance doublet of \ion{P}{v} ($\lambda\lambda$ 1118, 1128) is the only 
visible feature of phosphorus and is clearly stronger in \agk. The observed and 
modelled relative strength of both components do not match perfectly. This could 
be explained by the $\lambda$1128 line being blended with a strong \ion{Ni}{vi} 
line.

\paragraph{Sulfur}
The \ion{S}{vi} resonance doublet is prominent in the \textit{FUSE} spectrum 
while the other expected sulfur lines (\ion{S}{vi} $\lambda$1117.7 and 
\ion{S}{v} $\lambda\lambda$1122.0, 1128.6, 1133.9) are absent or weak. 
In the spectra of \agk\ and \lsii\, the \ion{S}{v} $\lambda$1122.0 is blended 
with an IS line of \ion{Fe}{ii}.

\paragraph{Iron}

The iron abundance was determined by
fitting the spectral region between 1360 and 1480 \AA, which is mostly
dominated by strong \ion{Fe}{v} lines. Fig.~\ref{iuefe} shows a comparison 
between the spectrum of \feige\ and our best model for this spectral region. The 
figure shows a very
good agreement between the observed and synthetic spectra. The
synthetic spectrum was calculated using the latest line
compilation\footnote{http://kurucz.harvard.edu/linelists, dated Dec. 2016, 
including lines computed only between known levels, thus with good 
wavelengths.} by Kurucz, which provides a much
larger number of lines than in the previous compilation \citep{kur95}. For 
instance,
the number of \ion{Fe}{v} lines went from 1336 to 5580 lines between 900 and
2000 \AA. Although the addition of these new iron lines has little effect on
the match between the model and the \textit{IUE} spectrum, it significantly 
improves the model in the \textit{FUSE} spectral range, especially for
$\lambda \leq 1050$ \AA\ (see Fig.~\ref{fuse}). Some of the new lines appear
stronger in the model than in the observations, however.

\paragraph{Nickel}
The majority of nickel lines are found at wavelengths shorter than 1320 \AA. 
We selected the \textit{FUSE} and \textit{IUE} ranges 1165-1185 \AA\ and 
1220-1265 \AA\ to determine the nickel abundance.  Fig.~\ref{iueni} compares our 
best model of \feige\ to its \textit{IUE} portion. The figure not only shows 
many \ion{Ni}{v} lines, but also several \ion{Fe}{vi} lines.

\paragraph{Chromium, manganese, and cobalt}
We looked for these elements in the \textit{FUSE} and \textit{IUE} spectra, but 
did not unambiguously detect lines.
We measured abundance upper limits by comparing the strongest predicted
and isolated lines to the observed spectra.
Because of the lack of model atoms, we computed the energy level populations based 
on a LTE
approximation, but used the resulting physical structure provided by
our final NLTE model atmosphere.

\paragraph{}
A comparison between the \textit{FUSE} spectrum of \feige\ and a synthetic one 
using the latest line list by Kurucz is presented in Fig.~\ref{fuse}.
The synthetic model includes the abundances listed in Table \ref{abund} and Cr, Mn, and Co at their upper limits. 
To identify IS absorption features in the \textit{FUSE} range we used the OWENS 
programme developed by the \textit{FUSE} French team \citep{hebr03}. The programme 
allows the user to model the IS absorption produced by different clouds of IS material with 
distinct radial and turbulent velocities, temperature, and column density of 
various elements. The resulting IS lines are indicated in green.

\section{Spectral energy distributions and distances}

\begin{figure*}
\begin{center}
\includegraphics[width=1\linewidth]{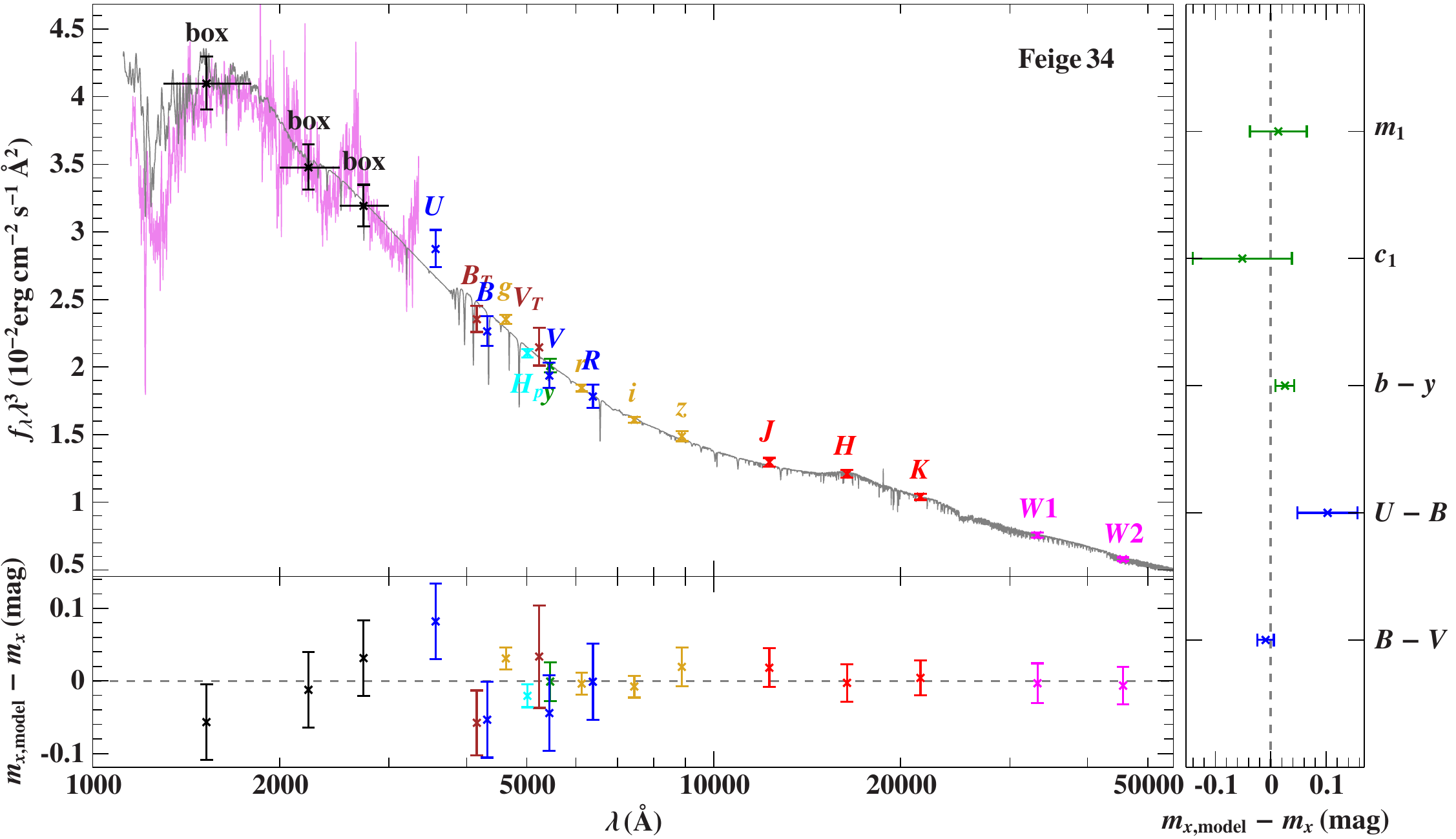}
\caption{Comparison of our best fitting synthetic and observed photometry for 
the \feige\ binary system.
The different photometric systems have the following colour code: Johnson-Cousins 
(blue), Strömgren (green), Tycho (brown), Hipparcos (cyan), SDSS (gold), 2MASS 
(red), WISE (magenta), and Geneva (crimson).
Theoretical reddened spectra are shown in grey and the \textit{IUE} data in 
pink.}
\label{photom}
\end{center}
\end{figure*}

\begin{table*}
\caption{Results of the spectral energy distribution fitting and derived distances}
\label{dist}
\centering
\begin{tabular}{lcccc}
\hline
\hline
   & \feige & \feiget  & \agk & \lsii    \\
\hline
$E(B-V)$ & 0.018 $\pm$ 0.007  &  0.031 $\pm$ 0.004 & 0.023 $\pm$0.003  &  0.060 
$\pm$ 0.005     \\
$\theta$ ($\times 10^{11}$) &  2.01 $\pm$ 0.03 & 1.533 $\pm$0.008 & 1.423 $\pm$ 
0.006  & 1.39 $\pm$ 0.01   \\
D$_{\rm spectro}$  (pc) & 257 $\pm$ 36  & 337 $\pm$ 44  & 335 $\pm$ 44  & 371 
$\pm$ 50   \\
D$_{\rm Hipparcos}$ (pc)\tablefootmark{a} & 323$^{+539}_{-124}$ & 
273$^{+446}_{-104}$   &  265$^{+293}_{-91}$   &  ...   \\
\hline
\teff\ $_{\rm comp}$ (K)  & 3848$^{+214}_{-309}$ & ...  & ...  & ...   
\\
log $g$ $_{\rm comp}$ & 5.4$^{+0.1}_{-1.3}$ & ...  & ...  & ...  \\
Surface ratio &  23.5 $\pm$ 2.0  & ...  & ...  & ... \\
    
\hline
\end{tabular}\\
\tablefoot{
\tablefoottext{a}{from \cite{hip07}}
}
\end{table*}

We collected photometry data from a variety of sources to construct the SED of our four stars ranging from the UV range up to the 
infrared ($\approx$ 4.6 $\mu$m). The photometric systems and catalogues used are 
listed in Table~\ref{photocat}.
The UV range is represented by low resolution, large aperture, flux calibrated 
\textit{IUE} spectra that were used to create three artificial magnitudes of 1300$-$1800 \AA, 2000$-$2500 \AA, and 2500$-$3000 \AA\  via 
box filters. The 
\textit{IUE} spectra with the longest exposure time for each star were retrieved 
from the MAST archive to create these UV magnitudes.
To model the reddening we assumed a standard value of the extinction parameter 
($R_{V}$ = 3.1) and used the \citet{fitz99} extinction law. Our best synthetic 
spectra were used to compute synthetic magnitudes. 

The fitting procedure aims at achieving the best match between synthetic and 
observed magnitudes and colours by $\chi^2$ minimization. For a single star fit, 
the angular diameter ($\theta$ = 2R/D) and colour excess $E(B-V)$ were the two 
free parameters. The resulting best fit for \feige\ is presented in 
Fig.~\ref{photom} and the corresponding parameters for all stars are listed in 
Table \ref{dist}, along with the spectroscopic and Hipparcos distances.  
The SEDs of \feiget, \agk, and \lsii\ are presented in Fig. \ref{photomadd}. 
They are well reproduced by our sdO models and do not show evidence of a 
companion. According to our radial velocity measurements, the stars also do not 
show any large RV variations; see Appendix A and Table \ref{rvbokblue} and 
\ref{rvlowres}.   

The situation is different for \feige,\ which requires a companion to 
reproduce the infrared excess. 
In this case, we used the library of PHOENIX \citep{haus99,haus99b} 
synthetic spectra computed by \citet{hus13} to model the photometry of the cool 
companion.
The SED of the composite spectra requires three additional parameters to be 
fitted; i.e. the \teff\ and log $g$ of the companion, and the surface ratio of both 
stars ($\propto(R_{\rm comp}/R_{\rm primary})^2$). 
Our best fit indicates \teff\ $\approx$ 3850 K for the companion, corresponding 
to a M0 spectral type. This is in good agreement with \citet{mull07} who found 
the infrared excess of \feige\ to be well reproduced by a stellar model at 3750 
K. They used photometry from the IRAC camera on the Spitzer Space Telescope 
combined with 2MASS and optical magnitudes to search for infrared excess in a 
sample of white dwarfs.\footnote{Because it is part of the \citet{mccook87} 
white dwarf catalogue, the star is at times included in surveys targeting bright 
white dwarfs.}

Using the angular diameter provided by the SED fitting we computed spectroscopic 
distances. 
The errors on D$_{\rm spectro}$ provided in Table \ref{dist} consider the 
uncertainties reported on $\theta$, the assumption of a canonical mass of 0.47 $\pm$ 0.01 \msun\ 
and uncertainties of $\pm$0.1 dex on the surface gravities. The errors provided 
must be seen as lower bounds, given that other sources of errors, such as uncertainties on \teff\,, the reddening, and larger 
deviations from the canonical mass, are likely
to 
contribute. 
Our spectroscopic distances are in agreement with the Hipparcos measurements 
within their (rather large) uncertainties. 
For \feiget\ and \agk, our D$_{\rm spectro}$ is nevertheless larger than D$_{\rm 
Hipparcos}$, while for \feige\ we have the opposite situation. According to 
Hipparcos, it should be the farthest star in our sample, but the closest star  
according to the spectroscopic distance. Most certainly Gaia will soon provide 
more accurate parallaxes for the three standard stars and for \lsii.

There have also been distance discrepancies reported in 
previous analyses of other hot sdOs (Feige 110, BD$+$75$\degr$325, and \bd; 
\citealt{rauch14,lanz97,lat15})
and a DAO-type white dwarf (LSV$+$46$\degr$21; \citealt{rauch07}). In these four 
cases, the spectroscopic distances derived were significantly larger than the 
Hipparcos values.


\section{Discussion}
\subsection{Evolutionary status}
\begin{figure}[t]
\begin{center}
\resizebox{\hsize}{!}{\includegraphics{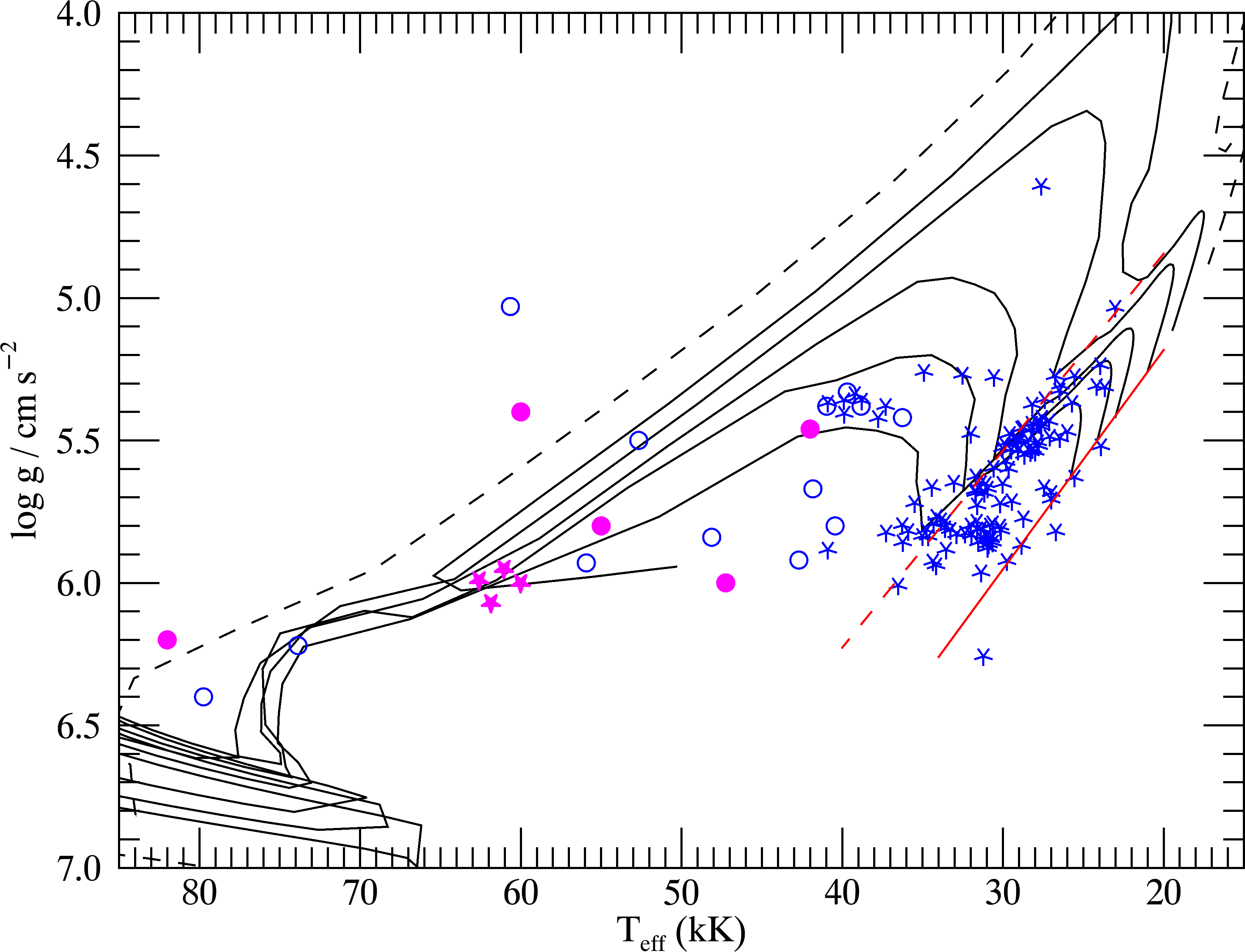}}
\caption{Position of our four stars (magenta stars) in the \teff~$-$log~$g$ 
diagram compared to other hydrogen-rich sdBs and sdOs and evolutionary 
tracks. The post-EHB tracks are from the Y=0.288 set of \citet{dor93} for total 
masses of 0.471, 0.473, 0.475, 0.480, and 0.485 \msun\ from left to right. An 
additional post-early AGB track is also illustrated (dashed 0.495 \msun). The 
red lines define the EHB region where core helium burning takes place. Observed 
data are from subsamples of H-rich stars taken from \citet[asterisk]{font14} and 
\citet[circle]{stro07}. Also included are the position of three other sdOs for which 
chemical abundances are known (AA Dor, Feige 110, and \bd) as well as EC 
11481$-$2303 and CPD$-$71$\degr$172B (magenta filled circles). }
\label{gteff}
\end{center}
\end{figure}

From an evolutionary point of view, H-rich sdOs are the natural progeny of the 
H-rich sdBs. Most of these are found in the post-EHB region of the 
\teff~$-$log~$g$ diagram; this is the He-shell burning phase that happens after 
the star reaches the terminal-age EHB (dashed red curve in Fig.~\ref{gteff}). 
The most luminous among the hot sdOs can be explained if they are slightly more 
massive than EHB stars and are now in a post-AGB or post-early AGB phase, during which 
they evolve at higher luminosity due to their higher mass; the dashed track 
in Fig.~\ref{gteff} is for a post-early AGB of 0.495 \msun.
The position of our four stars in the \teff~$-$log~$g$ diagram is in good 
agreement with these stars being in a post-EHB phase in the region where the 
evolutionary tracks converge to similar surface gravity values. In addition, the 
agreement between the spectroscopic and Hipparcos distances of our three 
brightest stars does not suggest masses significantly different from the 
canonical value of EHB stars.

\begin{figure*}
\includegraphics[width=1\linewidth]{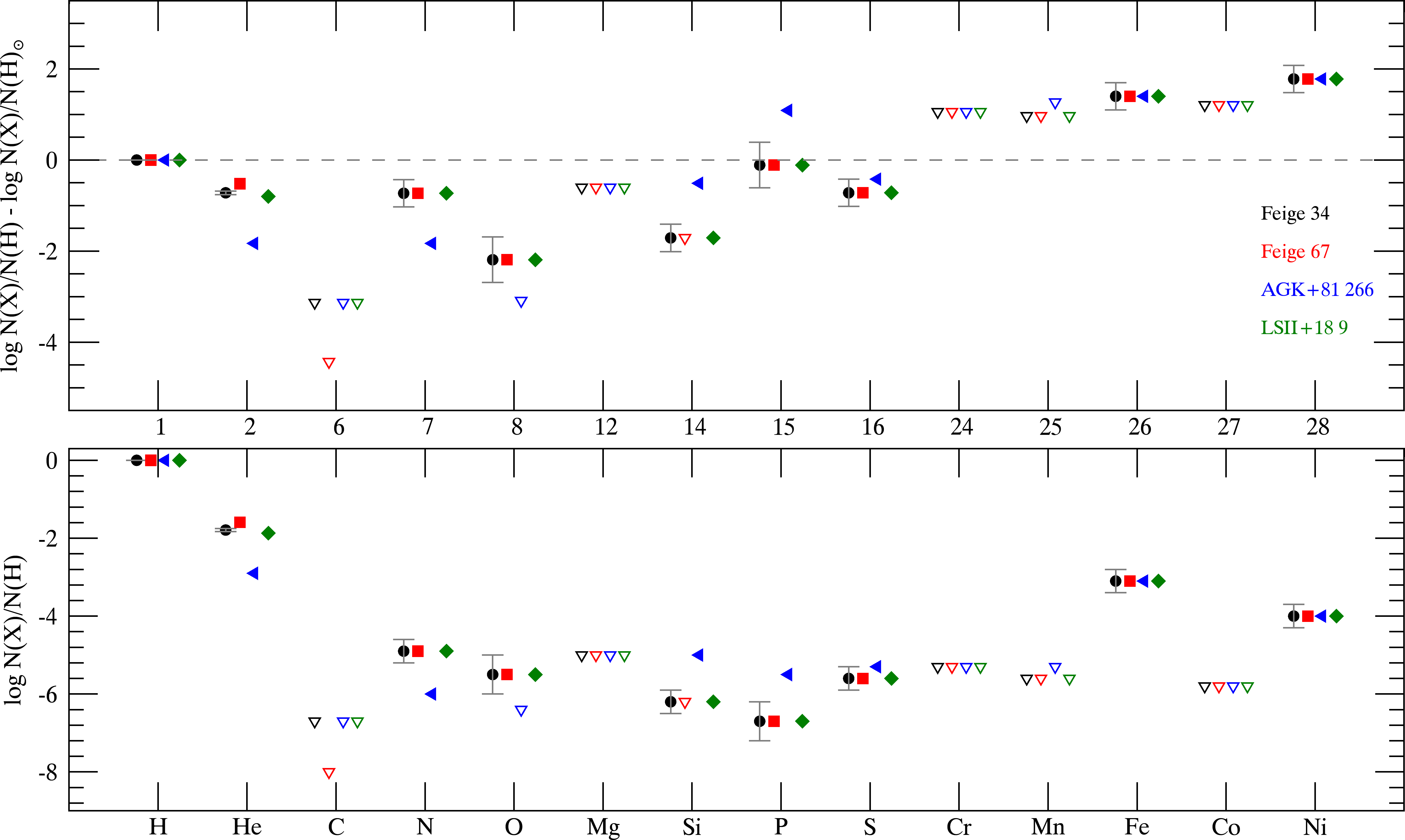}
   \caption{Summary of the chemical composition of \feige\ (black circle), 
\feiget\ (red square), \agk\ (blue left triangle), and \lsii\ (green diamond). 
Upper limits are represented by open downward triangles using the same colour 
code. Error bars are shown only for \feige\ but are similar for the other 
stars.  }
   \label{pabund}
\end{figure*}

\subsection{Diffusion processes}
Our four stars share almost identical atmospheric parameters and extremely 
similar spectra. Only \agk\ distinguishes itself with a helium abundance about 
10$ $ times lower than the three other stars. It also has a slightly different 
abundance pattern for the light elements even though it is equally enriched in iron and 
nickel. The derived abundances of our stars are summarized in Fig.~\ref{pabund}. 
Hot subdwarf stars are known to be chemically peculiar objects and their 
abundances are governed by a mixture of diffusion processes along with possible 
stellar winds and turbulence. The chemical composition of helium enriched 
subdwarfs can also be greatly affected by their evolutionary history. It goes without 
saying that star-to-star abundance variation is the norm among hot subdwarfs, 
although some general tendencies can be drawn 
\citep{bla08,otoole06,geier13,jef17}. 
Thus it seems rather unusual to have here a sample of four stars presenting 
almost the same atmospheric composition and especially strong enhancement in 
iron ($\approx$ 25$\times$ solar) and nickel ($\approx$ 60$\times$ solar). 

Another interesting star is 
CPD$-$71$\degr$172B. This star is the hot subdwarf component of a binary system in 
which the F-type star dominates in the optical spectral range. The hot component 
was only identified via the UV excess of the system and photometric fitting 
resulted in a \teff\ $\approx$ 60~000$\pm$5~000 K \citep{vit88}.
As mentioned previously, \citet{del92} reported the remarkable similarity of the 
\textit{IUE} spectra of CPD$-$71$\degr$172B with that of \lsii, which would make 
it also quite similar to the stars analysed in this work. We confirmed this 
statement by retrieving and comparing its \textit{FUSE} spectrum to those of our 
stars; it is equally similar to the spectra shown in Fig.~\ref{fusecomp}, suggesting 
comparable enrichment in iron and nickel. Small variations can be noticed in the 
lighter elements, for example \ion{He}{ii} $\lambda$1640 and the \ion{P}{v} 
doublet are similar to those of \agk. 
The surface gravity of CPD$-$71$\degr$172B was found to be lower than for our 
stars (log $g$ $\approx$ 5.4; \citealt{vit88}), however it was only 
estimated from the photometry of the binary system, so the actual value 
could well be higher. 

Another interesting object is the peculiar sdO star EC 11481$-$2303. Its UV 
spectrum indicates extreme heavy metal enhancement and it has an unusual 
UV continuum shape \citep{stys00,rauch10}. The atmospheric parameters (\teff\ = 
55~000$\pm$5000 K, log $g$ = 5.8$\pm$0.3) and hydrogen dominated atmosphere of this object
makes it relatively similar to our sdOs. Its position in the \teff~$-$log~$g$ 
plane is also in agreement with the star being in the post-EHB phase (see 
Fig.~\ref{gteff}).

Combining the results of our analysis of \feige, \feiget, \agk, and \lsii\ with 
the literature information on CPD$-$71$\degr$172B and EC 11481$-$2303, we are 
getting the picture of a rather homogeneous group of helium-poor post-EHB stars 
sharing similar atmospheric parameters and showing strong enrichment in iron and 
nickel. 
We could think that the stars have a combination of temperature and gravity that 
is favourable to iron and nickel enhancement but depletes most of the lighter 
elements and that the phenomenon causing their chemical pattern affects our 
four stars (and probably CPD$-$71$\degr$172B) in a very similar way. 
Given that the equilibrium radiative levitation theory stipulates that stars 
with
similar effective temperature and gravity should have similar abundances,
radiative levitation could play an essential role in maintaining the abundances 
in the atmospheres of the stars \citep{cha95}.
The computations carried out by \citet{rin12} for EC 11481$-$2303 indeed suggest 
that iron-peak elements are enhanced due to the effect of radiative levitation.

The He, N, Si, and P abundances observed in the atmosphere of \agk\
differ significantly from those observed in the other three stars
and suggest that other mechanisms could disrupt the abundances
supported by radiative levitation. Radiative levitation theory
predicts a unique abundance pattern for a set of effective temperature
and gravity and cannot explain the abundance discrepancies. Weak
stellar winds in competition with gravitational settling and radiative
levitation could affect the abundances. Indeed high effective
temperatures favour the presence of radiatively driven winds. \citet{krt16}
computed models for such winds and predicted mass loss
rates of 10$^{-10}$ and 10$^{-9}$ \msun\ yr$^{-1}$ for \feige\ and \feiget. 
Although
their atmospheric parameters are different than ours, their
calculations show that stellar winds could be present at the surface
of such stars. 
Stellar wind of such intensity have been observed previously in subdwarf 
stars, however with atmospheric parameters that are different than those of our sdOs 
\citep{lanz97,reindl14}. 
Nevertheless we recall here that 
no P Cygni profiles are detected in our spectra and weaker mass
loss rates could be invoked for disrupting the equilibrium abundances.

\subsection{Conservation of angular momentum}

The majority of sdB stars, with the exception of those found in close binary 
systems, are known to be slow rotators. In their study of rotational properties 
of 105 sdB stars, \citet{geier12} found that their
projected rotational velocities were lower than 10 km s$^{-1}$. In that case, 
it is surprising that our stars all show significantly higher rotation velocities. 
Similar high velocities have been previously reported by others.  For 
instance, \citet{beck95a} adopted a value of 30 km s$^{-1}$ for \feiget\ after 
inspection of the iron lines in its \textit{IUE} spectrum. As mentioned above, 
\citet{del92} noticed the presence of line broadening in the spectra of \lsii\ 
and CPD$-$71$\degr$172B, but they did not quantify it. The rounded cores 
observed in the \textit{FUSE} spectrum of CPD$-$71$\degr$172B confirms that it 
has a \vsini\ similar to that of our four objects. Interestingly, although the 
absorption features of EC 11481$-$2303 in the \textit{FUSE} range could not be 
very well modelled, \citet{rin12} reported that lines in the theoretical models 
were much narrower than in their observations; these authors estimated \vsini\ = 30 km s$^{-1}$. 

The observations of higher rotation velocities hint at  
conservation of angular momentum as these stars evolve away from the EHB. 
For instance, according to the evolutionary tracks of \citet{dor93}, the radius 
in the post-EHB region (close to the parameters of our stars) is about two-thirds
of the radius during the He-core burning phase. Conservation of angular 
momentum indicates that the rotation velocity would scale as 1/$R$ (assuming a 
solid homogeneous sphere), thus the rotation during the post-EHB phase should 
increase by $\approx$ 1.5. However, this is at least a factor two smaller than 
what is observed. Still, this simple calculation illustrates that conservation of 
angular momentum could be responsible for increasing the rotation velocity 
of a star as it evolves from the EHB. More detailed calculations taking into 
account the internal stellar structure will be needed to validate our 
hypothesis.

\subsection{ Feige 34 binary system}

Besides deriving spectroscopic distances, our SED fitting procedure allowed us 
to determine some parameters for the companion of \feige, whose temperature 
(3848$^{+214}_{-309}$ K) matches an early M or late K spectral type. 
Despite the clear presence of a companion, \feige\ does not show significant RV 
variations according to previous literature \citep{max00,han11}. We used our series 
of MMT spectra of \feige\ to compute radial velocities via cross-correlation of 
the different individual exposures as described in Appendix A. The resulting 
values are listed in Table \ref{rvf34}, where we find an average RV = 11.0 km 
s$^{-1}$ with a standard deviation of 7.7 km s$^{-1}$. Although we do 
not detect important RV variations in our data, there is some scatter of the 
individual measurements that appears to be significant (see discussion in 
Appendix A). Our radial velocity is higher than the value of $-$5.6$\pm$0.4 km 
s$^{-1}$ reported by \citet{max00} from WHT observations in 1993.  
\citet{max00} used the sharper emission peak in the H$_{\alpha}$ line to measure 
the radial velocity of \feige, whereas we were mainly limited to the cores
of three wide Balmer lines and the \ion{He}{ii} $\lambda$4686 line, since all
the metal lines are vanishingly small. However, very importantly, 
\citet{max00} obtained only 5 spectra over a period of 12 days, whereas our 
23 spectra were taken over many years.

A possibility that could explain the lack of RV variations is if the companion 
is a foreground or a background object. This can be tested by comparing the 
surface ratio derived during the SED fit with what can be expected for an sdO 
star with an M dwarf companion. The surface gravity of the companion cannot be 
well constrained by the photometric fit but the rather high value derived 
suggests a main-sequence star. Considering M = 0.47\msun\ and log $g$ = 6.0 for 
Feige 34 and a radius in the range 0.5$-$0.65 $R_{\sun}$ for the 
companion\footnote{for $\approx$ M2 to K7 main-sequence spectral type},
we obtained surface ratios between 20 and 32. These are only rough estimates but 
they are in agreement with the best fit value of 23.5, thus suggesting that both 
stars are at the same distance.

We are then left with the options of a rather long period or low inclination of 
the system (or a combination of both) to explain the lack of RV variations above 
$\approx$10 km s$^{-1}$. This would be in agreement with the photometric 
stability of the star (\citealt{marin16}) as a close M dwarf at high inclination 
would produce a reflection effect. 

\citet{lapal14} carried out the first systematic search for X-ray emission with 
the Chandra satellite for a sample of 19 bright ($V$ < 12) sdO stars. These authors 
detected X-rays in three of their targets, one of which was \feige\  with a 
detection above the 3$\sigma$ level. Given the known infrared excess of the 
star, the authors mention that the X-ray flux detected would be consistent with 
the expected range for an early M-type star. Thus they could not be certain of 
the origin of the observed X-rays. \feiget\ was also part of their sample but no 
detection was made for that source\footnote{identified as BD$+$18$\degr$2647 in 
their paper}. In the light of our results, knowing that both stars have very 
similar atmospheric parameters, we believe that the X-ray flux measured in 
\feige\ most likely originates from its M dwarf companion.

\section{Conclusions}

We have analysed \textit{FUSE}, \textit{IUE}, and optical spectra of four hot 
H-sdOs (\feige, \feiget, \agk, and \lsii) to derive their atmospheric 
parameters and chemical abundances.  
We used the high quality optical spectra to determine their atmospheric 
parameters and obtained very consistent values by fitting different spectra of a 
given star. This means that we were not limited by the quality, resolution, or 
spectral range of our observations. Our main source of errors on the atmospheric 
parameters would then come from systematic uncertainties related to the model 
atmospheres and would affect our four stars in the same way.
We confirm that \feige, \feiget, and \lsii\ are remarkably similar in terms of 
atmospheric parameters (\teff\ within 60\,000$-$63\,000 K, log $g$ 
within 5.9$-$6.1) and add to this group a fourth star, \agk. These parameters 
are in perfect agreement with a post-EHB nature for our stars. 

The UV spectra of the four stars are also extremely alike, indicating that in 
addition to a similarity in terms of atmospheric parameters, they also are similar in 
terms of chemical composition. They show an important enrichment in iron and 
nickel that is about 25 times and 60 times solar, respectively, while the 
other iron group elements examined (Cr, Mn, and Co) do not seem to have 
abundances higher than 10 times solar. The lighter elements are found to be 
solar or depleted. \agk\ distinguishes itself from the other three stars by 
slight differences in the abundances of its light elements, including helium. It 
is rather unexpected to have four (or three if we leave out \agk) stars with 
such similar abundances and spectra. Hot subdwarfs in general are known to show 
star-to-star variations in their chemical abundances and ``spectroscopic twins'' 
are definitely not common. It is not well understood how the different diffusion 
mechanisms govern the atmospheric abundances in sdBs and sdOs. In the particular 
case of our stars, radiative levitation could play an important role, as its 
effects are mainly determined by the temperature and surface gravity. Given the 
high effective temperature of our stars, stellar winds are predicted to be 
non-negligible, and we did not detect their typical P Cygni signature on 
resonance lines in any of our UV spectra.  The upper limits suggested for
the mass loss rates would be an interesting avenue to explore in 
future work.

We used photometric data to derive the spectroscopic distances of our stars and 
found our values to be in good agreement with the Hipparcos measurements 
available for our three brightest targets.  However, the current uncertainties on 
the parallax are rather large and Gaia will provide more accurate 
values very soon. Our photometric fit of \feige\ also puts additional constrains on 
the companion responsible for the infrared excess. Our results suggest a 
main-sequence star of about 3800 K and the surface ratio of the stars is 
consistent with these stars being at the same distance. However, according to our 
measurements \feige\ does not show radial velocity variations higher than 10 km 
s$^{-1}$, suggesting either a low inclination or a rather long orbital period. 

Another peculiar aspect in which our four stars are similar is 
their significant line broadening (\vsini\ $\approx$ 25 km s$^{-1}$). This appears 
to be at odds with their presumed post-EHB nature, as the sdB stars on the EHB 
are known to be notoriously slow rotators (\vsini\ < 10 km s$^{-1}$). A 
literature search highlighted two additional hot H-sdO stars 
(CPD$-$71$\degr$172B and EC 11481$-$2303) with atmospheric properties similar 
to our four stars and the spectra of both also indicate the presence of 
comparable rotational broadening. 
Computation of the angular momentum of stellar evolutionary models would be 
necessary to determine whether or not the rotation observed in our sdOs can be 
explained by angular momentum conservation as the stars contract during their 
post-EHB evolution.

\begin{acknowledgements}
This work was supported by a fellowship for postdoctoral
researchers from the Alexander von Humboldt Foundation awarded to M.L..
The computation of such an amount of model atmospheres was possible thanks to 
the CALYS cluster, which was funded by the Canada Research Chair of G. Fontaine 
and is maintained by P. Brassard.
We are also grateful to T. Rauch and U. Heber for interesting discussions and to 
S. Geier for observing time. 
Some of the data presented in this paper were obtained from the Mikulski Archive 
for Space Telescopes (MAST). STScI is operated by the Association of 
Universities for Research in Astronomy, Inc., under NASA contract NAS5-26555. 
Support for MAST for non-HST data is provided by the NASA Office of Space 
Science via grant NNX09AF08G and by other grants and contracts.
The MMT spectra analysed for this paper were obtained at the MMT 
Observatory, a joint facility of the University of Arizona and the Smithsonian 
Institution.
This research has made use of the SIMBAD and Vizier databases,
operated at CDS, Strasbourg, France.
This research makes use of the SAO/NASA Astrophysics 
Data System Bibliographic Service. 
This work used the profile-fitting procedure OWENS
developed by M. Lemoine and the  French Team.

\end{acknowledgements}

%
%

\bibliographystyle{aa}


\begin{thebibliography}{83}
\expandafter\ifx\csname natexlab\endcsname\relax\def\natexlab#1{#1}\fi

\bibitem[{{Ahn} {et~al.}(2012){Ahn}, {Alexandroff}, {Allende Prieto},
  {Anderson}, {Anderton}, {Andrews}, {Aubourg}, {Bailey}, {Balbinot}, {Barnes},
  \& et~al.}]{sdssdr9}
{Ahn}, C.~P., {Alexandroff}, R., {Allende Prieto}, C., {et~al.} 2012, \apjs,
  203, 21

\bibitem[{{Asplund} {et~al.}(2009){Asplund}, {Grevesse}, {Sauval}, \&
  {Scott}}]{asp09}
{Asplund}, M., {Grevesse}, N., {Sauval}, A.~J., \& {Scott}, P. 2009, \araa, 47,
  481

\bibitem[{{Bauer} \& {Husfeld}(1995)}]{bauer95}
{Bauer}, F. \& {Husfeld}, D. 1995, \aap, 300, 481

\bibitem[{{Becker} \& {Butler}(1995{\natexlab{a}})}]{beck95a}
{Becker}, S.~R. \& {Butler}, K. 1995{\natexlab{a}}, \aap, 294, 215

\bibitem[{{Becker} \& {Butler}(1995{\natexlab{b}})}]{beck95b}
{Becker}, S.~R. \& {Butler}, K. 1995{\natexlab{b}}, \aap, 301, 187

\bibitem[{{Becker} \& {Butler}(1995{\natexlab{c}})}]{beck95ni}
{Becker}, S.~R. \& {Butler}, K. 1995{\natexlab{c}}, \aap, 300, 453

\bibitem[{{Berger} \& {Fringant}(1978)}]{ber78}
{Berger}, J. \& {Fringant}, A.-M. 1978, \aap, 64, L9

\bibitem[{{Bergeron} {et~al.}(1992){Bergeron}, {Saffer}, \& {Liebert}}]{ber92}
{Bergeron}, P., {Saffer}, R.~A., \& {Liebert}, J. 1992, \apj, 394, 228

\bibitem[{{Bergeron} {et~al.}(1993){Bergeron}, {Wesemael}, {Lamontagne}, \&
  {Chayer}}]{ber93}
{Bergeron}, P., {Wesemael}, F., {Lamontagne}, R., \& {Chayer}, P. 1993, \apjl,
  407, L85

\bibitem[{{Blanchette} {et~al.}(2008){Blanchette}, {Chayer}, {Wesemael},
  {Fontaine}, {Fontaine}, {Dupuis}, {Kruk}, \& {Green}}]{bla08}
{Blanchette}, J.-P., {Chayer}, P., {Wesemael}, F., {et~al.} 2008, \apj, 678,
  1329

\bibitem[{{Bohlin} {et~al.}(2001){Bohlin}, {Dickinson}, \& {Calzetti}}]{boh01}
{Bohlin}, R.~C., {Dickinson}, M.~E., \& {Calzetti}, D. 2001, \aj, 122, 2118

\bibitem[{{Chayer} {et~al.}(1995){Chayer}, {Fontaine}, \& {Wesemael}}]{cha95}
{Chayer}, P., {Fontaine}, G., \& {Wesemael}, F. 1995, \apjs, 99, 189

\bibitem[{{Cutri}(2012)}]{cutri12}
{Cutri}, R.~M., e.~a. 2012, VizieR Online Data Catalog, 2311

\bibitem[{{D'Cruz} {et~al.}(1996){D'Cruz}, {Dorman}, {Rood}, \&
  {O'Connell}}]{dcruz96}
{D'Cruz}, N.~L., {Dorman}, B., {Rood}, R.~T., \& {O'Connell}, R.~W. 1996, \apj,
  466, 359

\bibitem[{{Deetjen}(2000)}]{deet00}
{Deetjen}, J.~L. 2000, \aap, 360, 281

\bibitem[{{Deleuil} \& {Viton}(1992)}]{del92}
{Deleuil}, M. \& {Viton}, M. 1992, \aap, 263, 190

\bibitem[{{Dorman} {et~al.}(1993){Dorman}, {Rood}, \& {O'Connell}}]{dor93}
{Dorman}, B., {Rood}, R.~T., \& {O'Connell}, R.~W. 1993, \apj, 419, 596

\bibitem[{{Drilling} \& {Heber}(1987)}]{dri87}
{Drilling}, J.~S. \& {Heber}, U. 1987, in IAU Colloq. 95: Second Conference on
  Faint Blue Stars, ed. A.~G.~D. {Philip}, D.~S. {Hayes}, \& J.~W. {Liebert},
  603--606

\bibitem[{{Fitzpatrick}(1999)}]{fitz99}
{Fitzpatrick}, E.~L. 1999, \pasp, 111, 63

\bibitem[{{Fleig} {et~al.}(2008){Fleig}, {Rauch}, {Werner}, \& {Kruk}}]{fle08}
{Fleig}, J., {Rauch}, T., {Werner}, K., \& {Kruk}, J.~W. 2008, \aap, 492, 565

\bibitem[{{Fontaine} {et~al.}(2014){Fontaine}, {Green}, {Brassard}, {Latour},
  \& {Chayer}}]{font14}
{Fontaine}, G., {Green}, E.~M., {Brassard}, P., {Latour}, M., \& {Chayer}, P.
  2014, in Astronomical Society of the Pacific Conference Series, Vol. 481, 6th
  Meeting on Hot Subdwarf Stars and Related Objects, ed. V.~{Van Grootel},
  E.~M. {Green}, G.~{Fontaine}, \& S.~{Charpinet}, 83

\bibitem[{{Geier}(2013)}]{geier13}
{Geier}, S. 2013, \aap, 549, A110

\bibitem[{{Geier} \& {Heber}(2012)}]{geier12}
{Geier}, S. \& {Heber}, U. 2012, \aap, 543, A149

\bibitem[{{Geier} {et~al.}(2015){Geier}, {Kupfer}, {Heber}, {Schaffenroth},
  {Barlow}, {{\O}stensen}, {O'Toole}, {Ziegerer}, {Heuser}, {Maxted},
  {G{\"a}nsicke}, {Marsh}, {Napiwotzki}, {Br{\"u}nner}, {Schindewolf}, \&
  {Niederhofer}}]{geier15}
{Geier}, S., {Kupfer}, T., {Heber}, U., {et~al.} 2015, \aap, 577, A26

\bibitem[{{Green} {et~al.}(2014){Green}, {Johnson}, {Wallace}, {O'Malley},
  {Amaya}, {Biddle}, \& {Fontaine}}]{gre14}
{Green}, E., {Johnson}, C., {Wallace}, S., {et~al.} 2014, in Astronomical
  Society of the Pacific Conference Series, Vol. 481, 6th Meeting on Hot
  Subdwarf Stars and Related Objects, ed. V.~{van Grootel}, E.~{Green},
  G.~{Fontaine}, \& S.~{Charpinet}, 161

\bibitem[{{Green} {et~al.}(2008){Green}, {Fontaine}, {Hyde}, {For}, \&
  {Chayer}}]{gre08}
{Green}, E.~M., {Fontaine}, G., {Hyde}, E.~A., {For}, B.-Q., \& {Chayer}, P.
  2008, in Astronomical Society of the Pacific Conference Series, Vol. 392, Hot
  Subdwarf Stars and Related Objects, ed. {U.~Heber, C.~S.~Jeffery, \&
  R.~Napiwotzki}, 75--+

\bibitem[{Haas(1997)}]{haas97}
Haas, S. 1997, PhD thesis, Friedrich-Alexander-Universit\"{a}t
  Erlangen-N\"{u}rnberg

\bibitem[{{Haas} {et~al.}(1996){Haas}, {Dreizler}, {Heber}, {Jeffery}, \&
  {Werner}}]{haas96}
{Haas}, S., {Dreizler}, S., {Heber}, U., {Jeffery}, S., \& {Werner}, K. 1996,
  \aap, 311, 669

\bibitem[{{Han} {et~al.}(2011){Han}, {Burlakova}, {Valyavin}, {Kim},
  {Galazutdinov}, {Zharikov}, {Lee}, {Kim}, {Kholtygin}, {Shulyak}, {Chavez},
  \& {Bertone}}]{han11}
{Han}, I., {Burlakova}, T., {Valyavin}, G., {et~al.} 2011, in Magnetic Stars,
  415--418

\bibitem[{{Han} {et~al.}(2003){Han}, {Podsiadlowski}, {Maxted}, \&
  {Marsh}}]{han03}
{Han}, Z., {Podsiadlowski}, P., {Maxted}, P.~F.~L., \& {Marsh}, T.~R. 2003,
  \mnras, 341, 669

\bibitem[{{Hauschildt} {et~al.}(1999){Hauschildt}, {Allard}, {Ferguson},
  {Baron}, \& {Alexander}}]{haus99b}
{Hauschildt}, P.~H., {Allard}, F., {Ferguson}, J., {Baron}, E., \& {Alexander},
  D.~R. 1999, \apj, 525, 871

\bibitem[{{Hauschildt} \& {Baron}(1999)}]{haus99}
{Hauschildt}, P.~H. \& {Baron}, E. 1999, Journal of Computational and Applied
  Mathematics, 109, 41

\bibitem[{{Heber}(2009)}]{heb09}
{Heber}, U. 2009, \araa, 47, 211

\bibitem[{{Heber}(2016)}]{heb16}
{Heber}, U. 2016, \pasp, 128, 082001

\bibitem[{{H{\'e}brard} \& {Moos}(2003)}]{hebr03}
{H{\'e}brard}, G. \& {Moos}, H.~W. 2003, \apj, 599, 297

\bibitem[{{Henden} {et~al.}(2015){Henden}, {Levine}, {Terrell}, \&
  {Welch}}]{apass}
{Henden}, A.~A., {Levine}, S., {Terrell}, D., \& {Welch}, D.~L. 2015, in
  American Astronomical Society Meeting Abstracts, Vol. 225, American
  Astronomical Society Meeting Abstracts, 336.16

\bibitem[{{H{\o}g} {et~al.}(2000){H{\o}g}, {Fabricius}, {Makarov}, {Urban},
  {Corbin}, {Wycoff}, {Bastian}, {Schwekendiek}, \& {Wicenec}}]{hog00}
{H{\o}g}, E., {Fabricius}, C., {Makarov}, V.~V., {et~al.} 2000, \aap, 355, L27

\bibitem[{{Husser} {et~al.}(2013){Husser}, {Wende-von Berg}, {Dreizler},
  {Homeier}, {Reiners}, {Barman}, \& {Hauschildt}}]{hus13}
{Husser}, T.-O., {Wende-von Berg}, S., {Dreizler}, S., {et~al.} 2013, \aap,
  553, A6

\bibitem[{{Jeffery} {et~al.}(2017){Jeffery}, {Baran}, {Behara}, {Kvammen},
  {Martin}, {Naslim}, {{\O}stensen}, {Preece}, {Reed}, {Telting}, \&
  {Woolf}}]{jef17}
{Jeffery}, C.~S., {Baran}, A.~S., {Behara}, N.~T., {et~al.} 2017, \mnras, 465,
  3101

\bibitem[{{Johnson} {et~al.}(2014){Johnson}, {Green}, {Wallace}, {O'Malley},
  {Amaya}, {Biddle}, \& {Fontaine}}]{john14}
{Johnson}, C., {Green}, E., {Wallace}, S., {et~al.} 2014, in Astronomical
  Society of the Pacific Conference Series, Vol. 481, 6th Meeting on Hot
  Subdwarf Stars and Related Objects, ed. V.~{van Grootel}, E.~{Green},
  G.~{Fontaine}, \& S.~{Charpinet}, 153

\bibitem[{{Klepp} \& {Rauch}(2011)}]{klepp11}
{Klepp}, S. \& {Rauch}, T. 2011, \aap, 531, L7

\bibitem[{{Krti{\v c}ka} {et~al.}(2016){Krti{\v c}ka}, {Kub{\'a}t}, \& {Krti{\v
  c}kov{\'a}}}]{krt16}
{Krti{\v c}ka}, J., {Kub{\'a}t}, J., \& {Krti{\v c}kov{\'a}}, I. 2016, \aap,
  593, A101

\bibitem[{{Kurucz} \& {Bell}(1995)}]{kur95}
{Kurucz}, R. \& {Bell}, B. 1995, Atomic Line Data (R.L.~Kurucz and B.~Bell)
  Kurucz CD-ROM No.~23.~Cambridge, Mass.: Smithsonian Astrophysical
  Observatory, 1995., 23

\bibitem[{{La Palombara} {et~al.}(2014){La Palombara}, {Esposito},
  {Mereghetti}, \& {Tiengo}}]{lapal14}
{La Palombara}, N., {Esposito}, P., {Mereghetti}, S., \& {Tiengo}, A. 2014,
  \aap, 566, A4

\bibitem[{{Landstreet} {et~al.}(2012){Landstreet}, {Bagnulo}, {Fossati},
  {Jordan}, \& {O'Toole}}]{land12}
{Landstreet}, J.~D., {Bagnulo}, S., {Fossati}, L., {Jordan}, S., \& {O'Toole},
  S.~J. 2012, \aap, 541, A100

\bibitem[{{Lanz} \& {Hubeny}(1995)}]{lanz95}
{Lanz}, T. \& {Hubeny}, I. 1995, \apj, 439, 905

\bibitem[{{Lanz} \& {Hubeny}(2003)}]{lanz03}
{Lanz}, T. \& {Hubeny}, I. 2003, \apjs, 146, 417

\bibitem[{{Lanz} {et~al.}(1997){Lanz}, {Hubeny}, \& {Heap}}]{lanz97}
{Lanz}, T., {Hubeny}, I., \& {Heap}, S.~R. 1997, \apj, 485, 843

\bibitem[{{Latour} {et~al.}(2013){Latour}, {Fontaine}, {Chayer}, \&
  {Brassard}}]{lat13}
{Latour}, M., {Fontaine}, G., {Chayer}, P., \& {Brassard}, P. 2013, \apj, 773,
  84

\bibitem[{{Latour} {et~al.}(2015){Latour}, {Fontaine}, {Green}, \&
  {Brassard}}]{lat15}
{Latour}, M., {Fontaine}, G., {Green}, E.~M., \& {Brassard}, P. 2015, \aap,
  579, A39

\bibitem[{{Marinoni} {et~al.}(2016){Marinoni}, {Pancino}, {Altavilla},
  {Bellazzini}, {Galleti}, {Tessicini}, {Valentini}, {Cocozza}, {Ragaini},
  {Braga}, {Bragaglia}, {Federici}, {Schuster}, {Carrasco}, {Castro},
  {Figueras}, \& {Jordi}}]{marin16}
{Marinoni}, S., {Pancino}, E., {Altavilla}, G., {et~al.} 2016, \mnras, 462,
  3616

\bibitem[{{Maxted} {et~al.}(2001){Maxted}, {Heber}, {Marsh}, \&
  {North}}]{max01}
{Maxted}, P.~f.~L., {Heber}, U., {Marsh}, T.~R., \& {North}, R.~C. 2001,
  \mnras, 326, 1391

\bibitem[{{Maxted} {et~al.}(2000){Maxted}, {Marsh}, \& {Moran}}]{max00}
{Maxted}, P.~F.~L., {Marsh}, T.~R., \& {Moran}, C.~K.~J. 2000, \mnras, 319, 305

\bibitem[{{McCook} \& {Sion}(1987)}]{mccook87}
{McCook}, G.~P. \& {Sion}, E.~M. 1987, \apjs, 65, 603

\bibitem[{{Mermilliod}(1997)}]{mer97}
{Mermilliod}, J.~C. 1997, VizieR Online Data Catalog, 2168

\bibitem[{{Miller Bertolami} {et~al.}(2008){Miller Bertolami}, {Althaus},
  {Unglaub}, \& {Weiss}}]{bert08}
{Miller Bertolami}, M.~M., {Althaus}, L.~G., {Unglaub}, K., \& {Weiss}, A.
  2008, \aap, 491, 253

\bibitem[{{Morel} \& {Magnenat}(1978)}]{mor78}
{Morel}, M. \& {Magnenat}, P. 1978, \aaps, 34, 477

\bibitem[{{Mullally} {et~al.}(2007){Mullally}, {Kilic}, {Reach}, {Kuchner},
  {von Hippel}, {Burrows}, \& {Winget}}]{mull07}
{Mullally}, F., {Kilic}, M., {Reach}, W.~T., {et~al.} 2007, \apjs, 171, 206

\bibitem[{{Napiwotzki}(2008)}]{nap08sdo}
{Napiwotzki}, R. 2008, in Astronomical Society of the Pacific Conference
  Series, Vol. 392, Hot Subdwarf Stars and Related Objects, ed. {U.~Heber,
  C.~S.~Jeffery, \& R.~Napiwotzki}, 139--+

\bibitem[{{Napiwotzki} {et~al.}(2004){Napiwotzki}, {Yungelson}, {Nelemans},
  {Marsh}, {Leibundgut}, {Renzini}, {Homeier}, {Koester}, {Moehler},
  {Christlieb}, {Reimers}, {Drechsel}, {Heber}, {Karl}, \& {Pauli}}]{nap04b}
{Napiwotzki}, R., {Yungelson}, L., {Nelemans}, G., {et~al.} 2004, in
  Astronomical Society of the Pacific Conference Series, Vol. 318,
  Spectroscopically and Spatially Resolving the Components of the Close Binary
  Stars, ed. R.~W. {Hilditch}, H.~{Hensberge}, \& K.~{Pavlovski}, 402--410

\bibitem[{{N{\'e}meth} {et~al.}(2012){N{\'e}meth}, {Kawka}, \&
  {Vennes}}]{nem12}
{N{\'e}meth}, P., {Kawka}, A., \& {Vennes}, S. 2012, \mnras, 427, 2180

\bibitem[{{Oke}(1990)}]{oke90}
{Oke}, J.~B. 1990, \aj, 99, 1621

\bibitem[{{O'Toole} \& {Heber}(2006)}]{otoole06}
{O'Toole}, S.~J. \& {Heber}, U. 2006, \aap, 452, 579

\bibitem[{{Pancino} {et~al.}(2012){Pancino}, {Altavilla}, {Marinoni},
  {Cocozza}, {Carrasco}, {Bellazzini}, {Bragaglia}, {Federici}, {Rossetti},
  {Cacciari}, {Balaguer N{\'u}{\~n}ez}, {Castro}, {Figueras}, {Fusi Pecci},
  {Galleti}, {Gebran}, {Jordi}, {Lardo}, {Masana}, {Mongui{\'o}},
  {Montegriffo}, {Ragaini}, {Schuster}, {Trager}, {Vilardell}, \&
  {Voss}}]{pan12}
{Pancino}, E., {Altavilla}, G., {Marinoni}, S., {et~al.} 2012, \mnras, 426,
  1767

\bibitem[{{Paunzen}(2015)}]{paun15}
{Paunzen}, E. 2015, \aap, 580, A23

\bibitem[{{Rauch} {et~al.}(2014){Rauch}, {Rudkowski}, {Kampka}, {Werner},
  {Kruk}, \& {Moehler}}]{rauch14}
{Rauch}, T., {Rudkowski}, A., {Kampka}, D., {et~al.} 2014, \aap, 566, A3

\bibitem[{{Rauch} {et~al.}(2010){Rauch}, {Werner}, \& {Kruk}}]{rauch10}
{Rauch}, T., {Werner}, K., \& {Kruk}, J.~W. 2010, \apss, 329, 133

\bibitem[{{Rauch} {et~al.}(2007){Rauch}, {Ziegler}, {Werner}, {Kruk},
  {Oliveira}, {Vande Putte}, {Mignani}, \& {Kerber}}]{rauch07}
{Rauch}, T., {Ziegler}, M., {Werner}, K., {et~al.} 2007, \aap, 470, 317

\bibitem[{{Reindl} {et~al.}(2014){Reindl}, {Rauch}, {Parthasarathy}, {Werner},
  {Kruk}, {Hamann}, {Sander}, \& {Todt}}]{reindl14}
{Reindl}, N., {Rauch}, T., {Parthasarathy}, M., {et~al.} 2014, \aap, 565, A40

\bibitem[{{Ringat} \& {Rauch}(2012)}]{rin12}
{Ringat}, E. \& {Rauch}, T. 2012, in Astronomical Society of the Pacific
  Conference Series, Vol. 452, Fifth Meeting on Hot Subdwarf Stars and Related
  Objects, ed. D.~{Kilkenny}, C.~S. {Jeffery}, \& C.~{Koen}, 71

\bibitem[{{Rufener}(1999)}]{geneva}
{Rufener}, F. 1999, VizieR Online Data Catalog, 2169

\bibitem[{{Sch\"{o}nberner} \& {Drilling}(1984)}]{schon84}
{Sch\"{o}nberner}, D. \& {Drilling}, J.~S. 1984, \apj, 278, 702

\bibitem[{{Stroeer} {et~al.}(2007){Stroeer}, {Heber}, {Lisker}, {Napiwotzki},
  {Dreizler}, {Christlieb}, \& {Reimers}}]{stro07}
{Stroeer}, A., {Heber}, U., {Lisker}, T., {et~al.} 2007, \aap, 462, 269

\bibitem[{{Stys} {et~al.}(2000){Stys}, {Slevinsky}, {Sion}, {Saffer},
  {Holberg}, {O'Donoghue}, {Kilkenny}, {Stobie}, \& {Koen}}]{stys00}
{Stys}, D., {Slevinsky}, R., {Sion}, E.~M., {et~al.} 2000, \pasp, 112, 354

\bibitem[{{Thejll} {et~al.}(1991){Thejll}, {MacDonald}, \& {Saffer}}]{the91}
{Thejll}, P., {MacDonald}, J., \& {Saffer}, R. 1991, \aap, 248, 448

\bibitem[{{Thejll} {et~al.}(1995){Thejll}, {Ulla}, \& {MacDonald}}]{the95}
{Thejll}, P., {Ulla}, A., \& {MacDonald}, J. 1995, \aap, 303, 773

\bibitem[{{Turnshek} {et~al.}(1990){Turnshek}, {Bohlin}, {Williamson}, {Lupie},
  {Koornneef}, \& {Morgan}}]{turn90}
{Turnshek}, D.~A., {Bohlin}, R.~C., {Williamson}, II, R.~L., {et~al.} 1990,
  \aj, 99, 1243

\bibitem[{{Valyavin} {et~al.}(2006){Valyavin}, {Bagnulo}, {Fabrika},
  {Reisenegger}, {Wade}, {Han}, \& {Monin}}]{val06}
{Valyavin}, G., {Bagnulo}, S., {Fabrika}, S., {et~al.} 2006, \apj, 648, 559

\bibitem[{{van Leeuwen}(2007)}]{hip07}
{van Leeuwen}, F. 2007, \aap, 474, 653

\bibitem[{{Viton} {et~al.}(1988){Viton}, {Burgarella}, {Cassatella}, \&
  {Prevot}}]{vit88}
{Viton}, M., {Burgarella}, D., {Cassatella}, A., \& {Prevot}, L. 1988, \aap,
  205, 147

\bibitem[{{Werner}(1996)}]{wer96}
{Werner}, K. 1996, \apjl, 457, L39

\bibitem[{{Werner} {et~al.}(1998){Werner}, {Dreizler}, {Haas}, \&
  {Heber}}]{wer98}
{Werner}, K., {Dreizler}, S., {Haas}, S., \& {Heber}, U. 1998, in ESA Special
  Publication, Vol. 413, Ultraviolet Astrophysics Beyond the IUE Final Archive,
  ed. W.~{Wamsteker}, R.~{Gonzalez Riestra}, \& B.~{Harris}, 301

\bibitem[{{Zhang} \& {Jeffery}(2012)}]{zhang12}
{Zhang}, X. \& {Jeffery}, C.~S. 2012, \mnras, 419, 452

\end{thebibliography}


\begin{appendix} 



\section{Radial velocity measurements}

Most of our observed optical spectra (MMT as well as low, 8.7 \AA, and 
medium, 1.3 \AA, resolution Bok spectra) consist of multiple exposures, taken with
the identical spectroscopic set-up
over a period of time ranging from one month to several years, which were 
combined together to obtain a final high sensitivity spectrum.  
The combination procedure consists of removing the continuum and cross-correlating 
(with the IRAF task \textit{fxcor}) each individual spectrum against a self-template 
derived from all of the spectra in an iterative process, as described in Sect.~3.1.
A Fourier filter was applied to restrict the cross-correlation to features narrower
than the half-width of the Balmer lines and wider than the instrumental resolution.
The advantage of cross-correlating the entire spectrum, except for the  small regions 
around interstellar lines that were excluded from the cross-correlations, against
a high S/N version of the same spectrum is that all of the information in each
spectrum contributes to the velocity precision; this includes weaker He and metal lines
that are partially hidden in the noise of the individual spectra.
As a by-product, the relative radial velocities returned 
by the cross-correlation procedure can be used to look for RV variations and to 
derive their absolute values. The latter were retrieved by measuring the RV of the 
combined spectrum and adding this value to each individual relative RV. The 
resulting values are reported in the last column of Tables \ref{rvf34} to 
\ref{rvlowres}, where the uncertainties are the sum (in quadrature) of the 
errors on the relative RV and the RV of the combined spectrum. The final RV 
of the star is obtained by computing the weighted average of the individual 
values and considering the standard deviation to be the uncertainty. 

Spectral resolution is a an important factor determining the accuracy of 
the radial velocity measurements, so we selected the MMT spectra (1.0 \AA) for 
\feige\ and the Bok (1.3 \AA) spectra for the three other stars. Since \agk\ has 
been observed only four times with the latter setting, we additionally retrieved
the RVs of the five lower resolution Bok (8.7 \AA) spectra of that star. The 
absolute radial velocity of the combined spectrum was measured by fitting a set of 
mathematical functions (Gaussian, Lorentzian, and polynomial) to the spectral 
lines using the FITSB2 routine \citep{nap04b}. These three functions can 
reproduce the line shape and continuum slope. The spectral lines are fitted 
simultaneously using $\chi^{2}$-minimization to determine the RV shift and 1 $\sigma$ 
error. We used  the H$\beta$, H$\gamma$, H$\delta$, 
and \ion{He}{ii} $\lambda$4686 lines for the MMT spectra, the 
Balmer lines from H$\gamma$ to H9 for the Bok 1.3
\AA\ , and finally  H$\alpha$ up 
to H8 and the \ion{He}{ii} $\lambda\lambda$4686, 5412 lines for the Bok 8.7 \AA. 
The relative RV standard deviations obtained with the Bok 1.3 \AA\ (6$-$7 km 
s$^{-1}$) are higher than the typical $\sigma _{RV}$ of 4$-$6 km s$^{-1}$ 
obtained for constant sdB stars using the same observational setting. However, 
sdB stars normally show, in addition to the Balmer lines, sharper \ion{He}{i} 
and metallic lines in the 3675$-$4520 \AA\ range, thus providing better 
accuracy for the cross-correlation procedure, whereas only six wide Balmer lines 
are available for the hot sdO stars.

\begin{table}[ht]
\caption{Radial velocities measured from the MMT spectra of \feige.}
\label{rvf34}
\begin{tabular}{lcccr}
\hline
\hline
 UT date & UT time & HJD & SNR & RV   \\
         &         &   ($-$2450000) &     & (km s$^{-1}$) \\
\hline
1996-03-11 & 03:20:54 &  153.64385 & 148.5 &  17.9 $\pm$  3.0 \\
1996-03-11 & 07:08:06 &  153.80163 & 201.4 &  19.5 $\pm$  2.9 \\
1996-12-17 & 11:21:39 &  434.97593 & 198.1 &  13.1 $\pm$  2.6 \\
1997-01-02 & 13:21:16 &  451.06005 & 208.7 &  13.8 $\pm$  2.5 \\
1997-01-28 & 08:52:44 &  476.87462 & 148.0 &  14.4 $\pm$  3.3 \\
1997-03-02 & 08:41:47 &  509.86697 & 170.3 &  10.8 $\pm$  2.9 \\
1997-03-03 & 06:02:37 &  510.75642 & 172.7 &  14.9 $\pm$  2.6 \\
1998-01-23 & 11:24:17 &  836.97972 & 176.2 &   8.7 $\pm$  2.7 \\
2002-01-23 & 05:46:06 & 2297.74486 & 235.8 &  10.4 $\pm$  2.2 \\
2002-01-23 & 06:35:25 & 2297.77911 & 359.7 &  12.1 $\pm$  2.0 \\
2002-01-23 & 08:24:33 & 2297.85490 & 346.8 &  10.1 $\pm$  2.0 \\
2002-04-24 & 04:29:44 & 2388.68900 & 488.1 &  26.3 $\pm$  2.0 \\
2002-04-24 & 07:16:44 & 2388.80497 & 120.1 &  24.1 $\pm$  3.8 \\
2002-04-24 & 07:32:09 & 2388.81567 & 113.1 &  25.2 $\pm$  3.4 \\
2004-12-31 & 11:41:20 & 3370.99053 & 276.6 &  11.4 $\pm$  2.2 \\
2010-03-24 & 08:50:24 & 5279.87210 & 263.4 &   1.1 $\pm$  2.2 \\
2010-06-11 & 03:52:37 & 5358.65934 & 390.1 &  11.6 $\pm$  2.6 \\
2010-12-13 & 12:39:22 & 5544.02957 & 283.5 &   7.2 $\pm$  2.5 \\
2010-12-13 & 12:43:02 & 5544.03211 & 349.7 &   5.2 $\pm$  2.3 \\
2011-01-13 & 12:01:51 & 5575.00542 & 299.5 &   4.4 $\pm$  2.4 \\
2013-01-03 & 11:52:18 & 6295.99831 & 221.3 &   9.5 $\pm$  2.5 \\
2013-01-03 & 12:13:00 & 6296.01269 & 215.4 &$-$6.4 $\pm$  2.4 \\
2013-01-03 & 12:21:04 & 6296.01829 & 151.0 &   4.1 $\pm$  3.1 \\

\hline
\multicolumn{3}{c}{Average}   & 1239 & 11.0 $\pm$ 7.7 \\
\hline
\end{tabular}
\end{table}

While it is clear from the results listed in Tables 
\ref{rvf34}$-$\ref{rvlowres} that there are no high RV variations in these stars, 
one still notices small fluctuations, especially in the cases of \feige\ and 
\lsii,\ where the standard deviation is larger than the average uncertainties. In 
order to estimate whether the RV variations could be statistical or not we 
computed the probability $p$ of obtaining such a sample from random fluctuations 
of constant value. We followed the method described in \cite{max01}, where the 
$\chi^2$ value is defined as 
   \begin{equation}
     \chi^2 = \sum_{i=1}^{N}\frac{(x_{i}-\bar{x})^2}{\Delta{x_i}^2}
   ,\end{equation}
where $x$ is the RV, $\bar{x}$ the average RV, and $\Delta{x_i}$ the error on 
the RV. The $\chi^2$ obtained is then compared with the $\chi^2$-distribution 
for the appropriate degree of freedom to obtain the probability $p$. From that 
test, we found that only the MMT data of \feige\ returned a probability smaller 
than 0.01\% (or log $p$ $\la$ $-$4.0), meaning that the variations in the data 
are significant. Despite that, we could not find a distinct period matching the 
data with a simple periodogram analysis, i.e. looking for sinusoidal 
variations.  This is very likely due, at least in part, to the sparse time sampling, 
which was a result of the fact that these spectra were obtained primarily for 
flux calibration rather than RVs.

\begin{table}[h]
\caption{Radial velocities measured from the Bok 1.3 \AA\ spectra.}
\label{rvbokblue}
\begin{tabular}{lcccl}
\hline
\hline
 UT date & UT time & HJD & SNR & RV   \\
         &         &  ($-$2450000) &     & (km s$^{-1}$) \\
\hline
\multicolumn{5}{l}{Feige 67} \\
2013-01-02 & 11:33:45 & 6294.98486 & 124.8 &  33.3 $\pm$  6.3 \\
2013-01-05 & 11:23:39 & 6297.97754 & 115.7 &  19.0 $\pm$  7.2 \\
2013-01-21 & 12:10:41 & 6314.01123 & 115.1 &  33.8 $\pm$  6.4 \\
2013-01-21 & 12:14:59 & 6314.01465 & 114.6 &  36.7 $\pm$  6.1 \\
2013-01-22 & 10:39:57 & 6314.94824 & 109.3 &  27.9 $\pm$  6.8 \\
2013-01-22 & 10:43:50 & 6314.95117 & 110.6 &  37.8 $\pm$  6.1 \\
2013-01-31 & 10:56:58 & 6323.96094 & 105.2 &  21.9 $\pm$  6.0 \\
2013-02-27 & 09:43:24 & 6350.91113 &  86.0 &  16.8 $\pm$  8.8 \\
2013-03-01 & 11:25:50 & 6352.98291 & 108.4 &  23.9 $\pm$  7.2 \\
\hline
\multicolumn{3}{c}{Average} &  331 & 27.0 $\pm$ 7.8 \\
\hline

\multicolumn{5}{l}{\lsii} \\
2012-06-21 & 09:13:04 & 6099.89062 & 132.2 & $-$43.2 $\pm$  6.3 \\
2012-06-21 & 09:20:44 & 6099.89600 & 127.9 & $-$39.6 $\pm$  6.1 \\
2012-06-21 & 11:04:52 & 6099.96826 & 114.6 & $-$37.7 $\pm$  8.6 \\
2012-08-01 & 07:59:10 & 6140.84033 & 103.5 & $-$56.1 $\pm$  7.7 \\
2012-08-01 & 08:08:04 & 6140.84668 & 106.7 & $-$57.2 $\pm$  7.5 \\
2012-08-02 & 07:46:07 & 6141.83105 & 117.0 & $-$65.1 $\pm$  6.5 \\
2012-08-02 & 07:54:37 & 6141.83691 & 113.5 & $-$62.4 $\pm$  6.8 \\
2012-08-03 & 08:46:12 & 6142.87256 & 112.9 & $-$47.5 $\pm$  7.8 \\
2012-08-03 & 08:54:42 & 6142.87891 & 113.2 & $-$58.0 $\pm$  6.8 \\
2012-08-29 & 05:29:43 & 6168.73584 & 138.0 & $-$27.4 $\pm$  6.8 \\
\hline
\multicolumn{3}{c}{Average} & 374 & $-$49.5 $\pm$ 12.3 \\
\hline

\multicolumn{5}{l}{\agk} \\
2013-01-05 & 09:37:59 & 6297.90625 & 116.2 & $-$33.6 $\pm$  6.3 \\
2013-01-22 & 08:34:03 & 6314.86133 & 114.7 & $-$12.4 $\pm$  5.4 \\
2013-01-22 & 08:39:31 & 6314.86523 & 115.9 & $-$31.2 $\pm$  6.1 \\
2013-01-31 & 08:16:53 & 6323.84912 &  78.7 & $-$15.8 $\pm$  5.8 \\
\hline
\multicolumn{3}{c}{Average} & 215 & $-$22.1 $\pm$ 10.7 \\

\hline
\end{tabular}
\end{table}

\begin{table}
\caption{Radial velocities measured from the Bok 8.7 \AA\ spectra of \agk.}
\label{rvlowres}
\begin{tabular}{lcccl}
\hline
\hline
 UT date & UT time & HJD & SNR & RV   \\
         &         &  ($-$2450000) &     & (km s$^{-1}$) \\
\hline
2003-12-01 & 12:14:46 & 2975.01367 & 164.7 & $-$34.5 $\pm$ 12.3 \\
2005-04-17 & 02:52:31 & 3477.62036 & 138.3 & $-$25.1 $\pm$ 13.5 \\
2006-02-12 & 07:37:43 & 3778.82129 & 168.7 & $-$19.4 $\pm$ 13.3 \\
2006-12-11 & 11:47:14 & 4080.99438 & 160.1 & $-$49.0 $\pm$ 12.7 \\
2010-03-28 & 04:18:38 & 5283.68066 & 154.2 & $-$30.6 $\pm$ 13.9 \\
\hline
\multicolumn{3}{c}{Average} & 352 & $-$34.1 $\pm$ 11.2 \\

\hline
\end{tabular}
\end{table}

Our final RV value for \feiget\ (27$\pm$8 km s$^{-1}$) is in good agreement with 
the value reported by \citet{dri87} of 29$\pm$6 km s$^{-1}$. As for \lsii, the 
same authors reported a RV of $-$25 km s$^{-1}$ (no errors) while the value 
measured by \citet{del92} is $-$50 km s$^{-1}$ (also no errors provided). Our 
own value of $-$49.5$\pm$12 km s$^{-1}$ is in better agreement with the latter.
In the case of \agk, we additionally retrieved the RVs of the five lower 
resolution Bok (8.7 \AA) spectra (Table \ref{rvlowres}). The mean RVs obtained 
with both sets of data agree within their uncertainties and we computed a final 
RV of $-$27.8$\pm$7.8 km s$^{-1}$. 

In conclusion, we do not detect any large variations in the radial 
velocities of our stars, but we cannot rule out smaller scale (below $\approx$10 
km s$^{-1}$) variations since some scatter in the individual measurements is 
seen. In the case of \feige\, a false-alarm probability test indicated that the 
variations cannot be purely statistical.  We cannot rule out that some 
instrumental effects could be unaccounted for in our evaluation of the 
individual uncertainties.  Such instrumental effects will be better characterized 
once the data for the whole sample of subdwarfs observed for this survey are 
processed. Nevertheless, with the exception of the
second from the last \feige\ RV value (for which we cannot point to any problem 
in the data), it is noteworthy that spectra obtained at nearly the same time all
result in very similar RVs, even as the average values change by a considerably
larger amount over longer time periods.  Furthermore, \citet{geier15} reported intriguing irregular RV variations in some He-rich sdO 
and \citet{gre14} observed irregular and occasional luminosity drops in some sdO 
light curves. Both phenomena still await an explanation. It could be that some 
similar, low scale variations are occurring among the stars of our sample.


\section{Additional material}

\begin{figure}[h]
\begin{center}
\resizebox{\hsize}{!}{\includegraphics{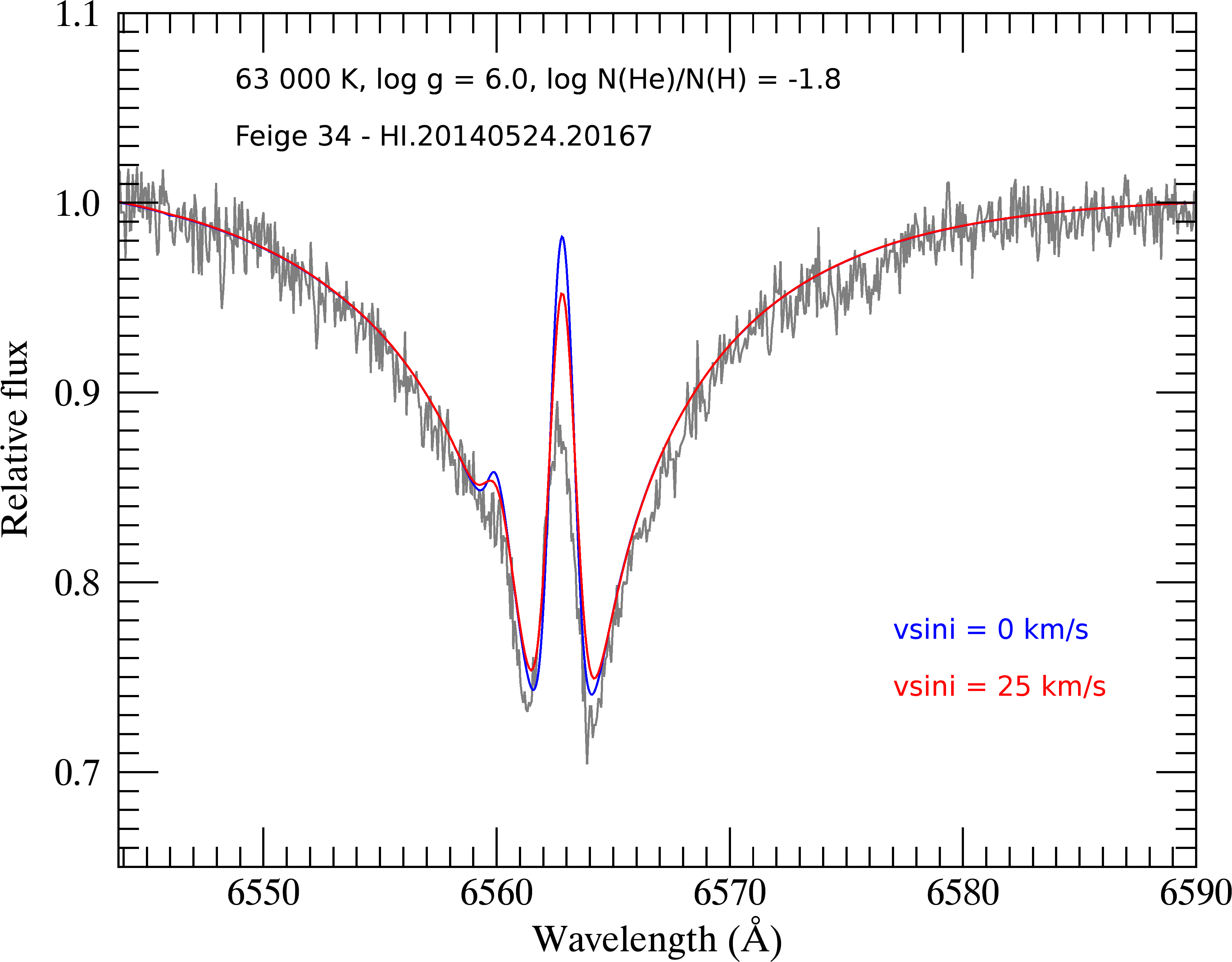}}
\caption{Comparison between a Keck HIRES spectrum of \feige\ and synthetic 
spectra with and without rotational broadening. }
\label{halpha}
\end{center}
\end{figure}


\begin{figure*}[p]
\begin{center}
\resizebox{\hsize}{!}{\includegraphics[angle=90]{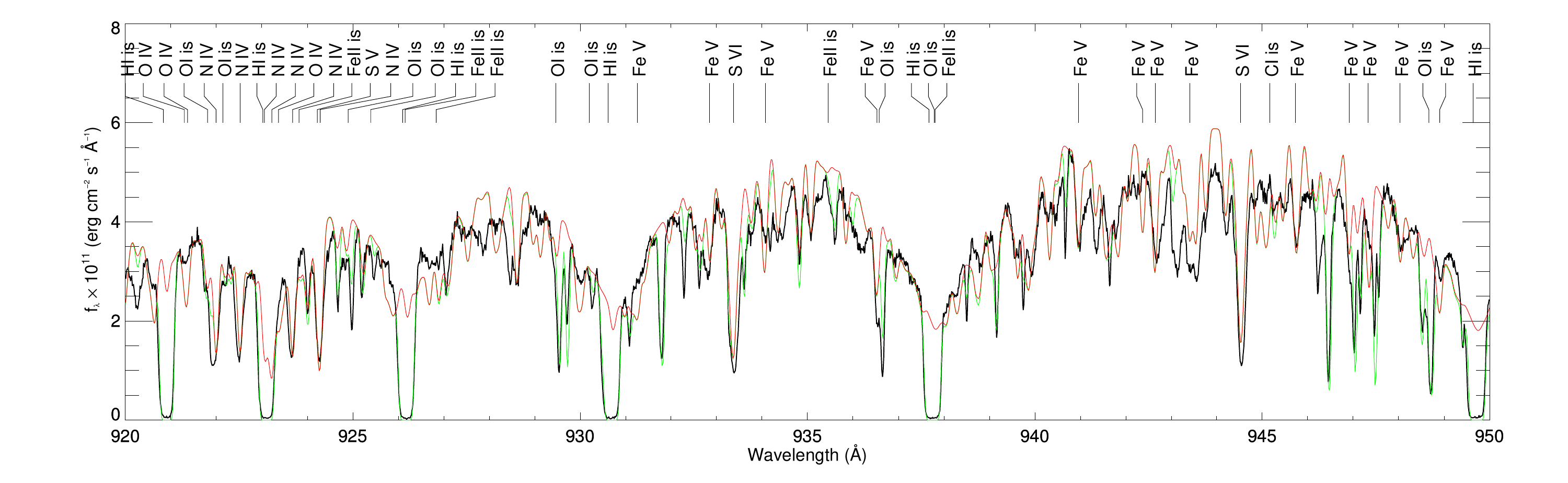}}
\resizebox{\hsize}{!}{\includegraphics[angle=90]{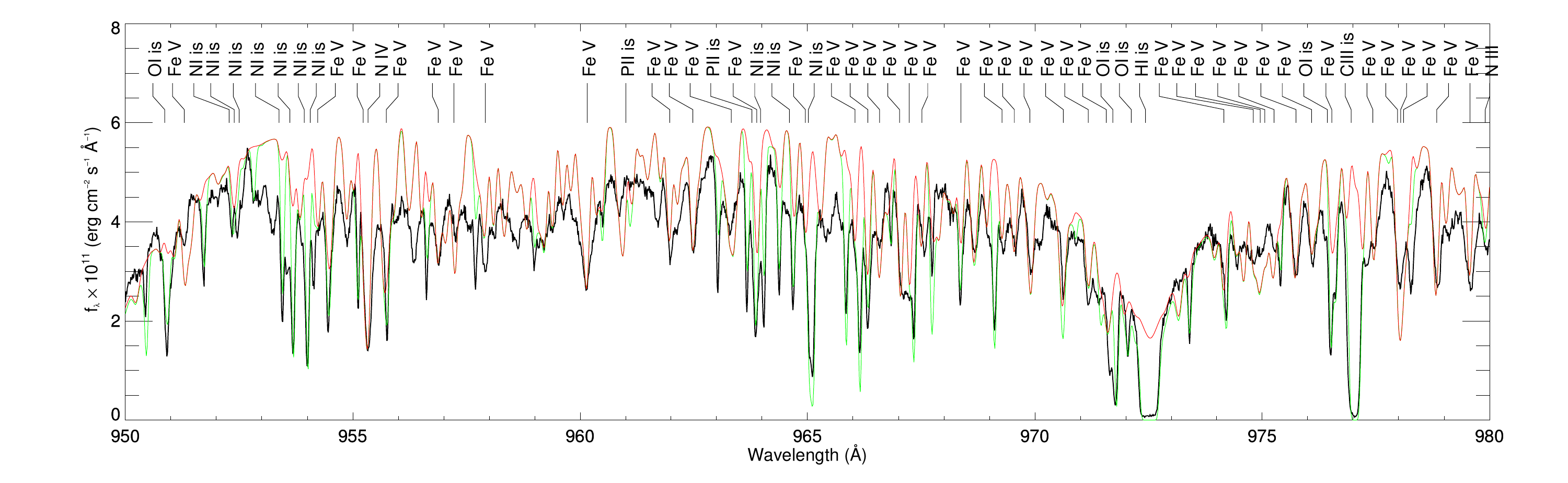}}
\resizebox{\hsize}{!}{\includegraphics[angle=90]{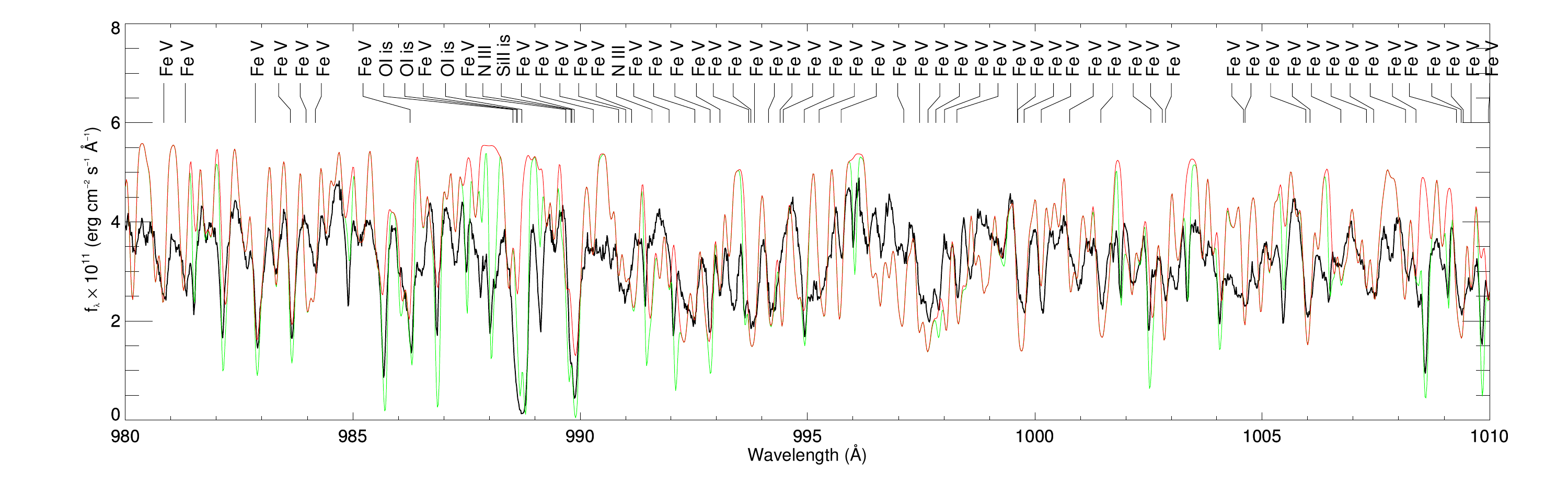}}
\resizebox{\hsize}{!}{\includegraphics[angle=90]{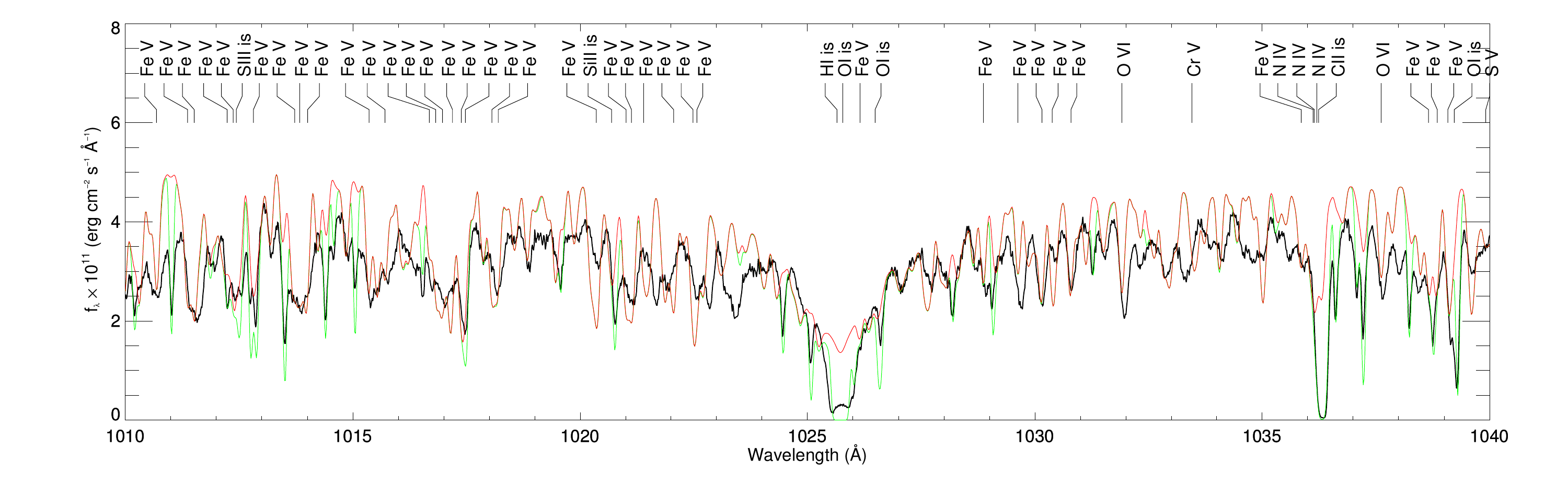}}

\caption{Comparison between the \textit{FUSE} spectrum (black) of 
\feige\ and a model spectrum using the latest Kurucz line list (red). The 
strongest lines are indicated at the top, as well as the IS lines included in 
the IS spectra (green). The IS lines without labels originate from H$_{2}$.}
\label{fuse}
\end{center}
\end{figure*}

\addtocounter{figure}{-1}
\begin{figure*}[p]
\begin{center}
\resizebox{\hsize}{!}{\includegraphics[angle=90]{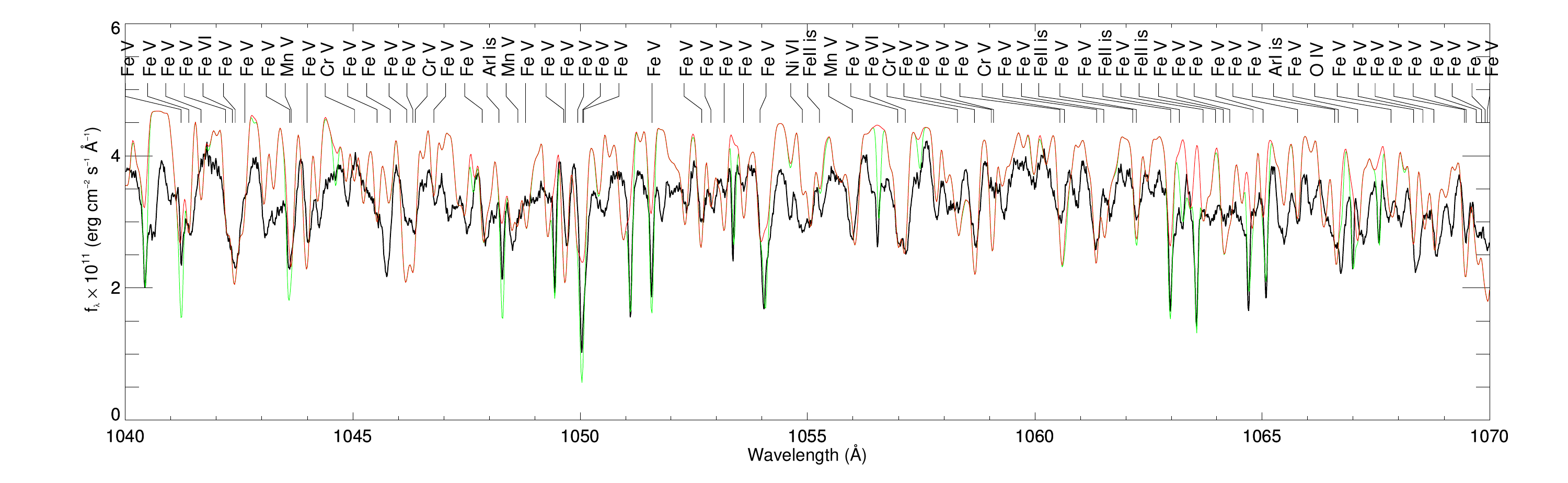}}
\resizebox{\hsize}{!}{\includegraphics[angle=90]{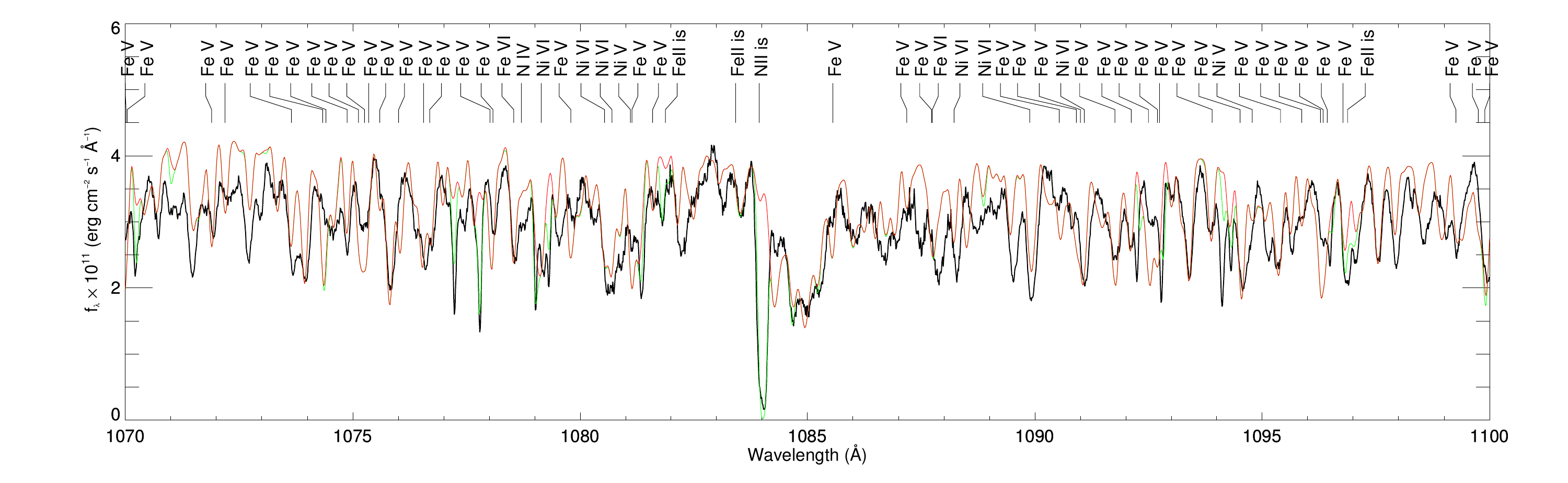}}
\resizebox{\hsize}{!}{\includegraphics[angle=90]{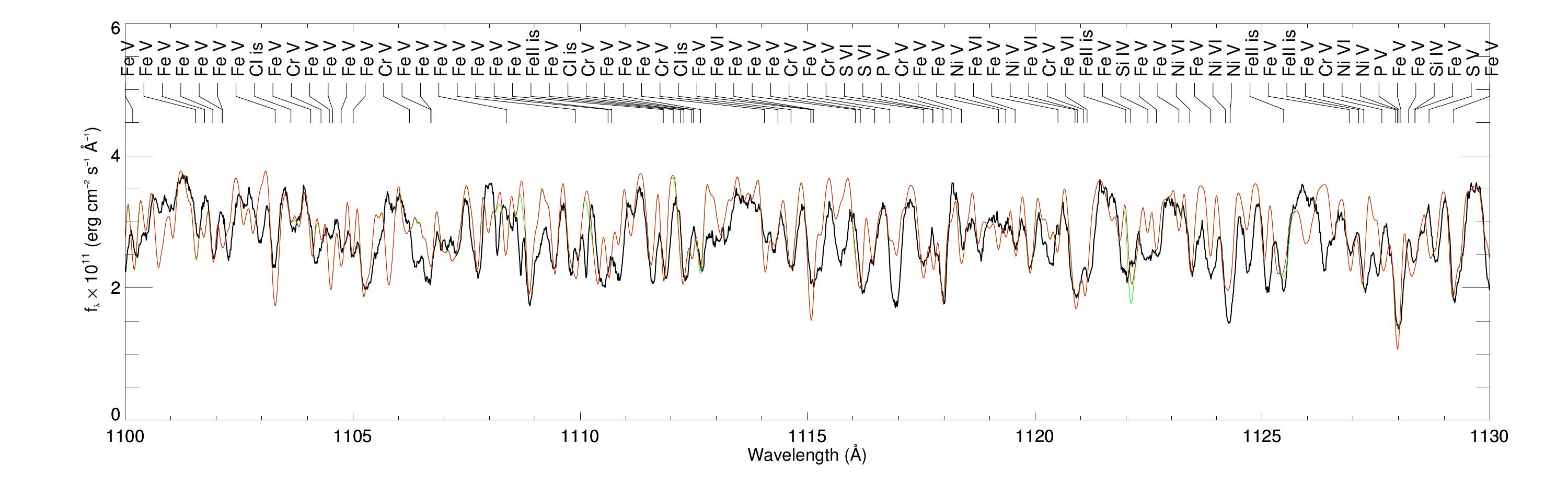}}
\resizebox{\hsize}{!}{\includegraphics[angle=90]{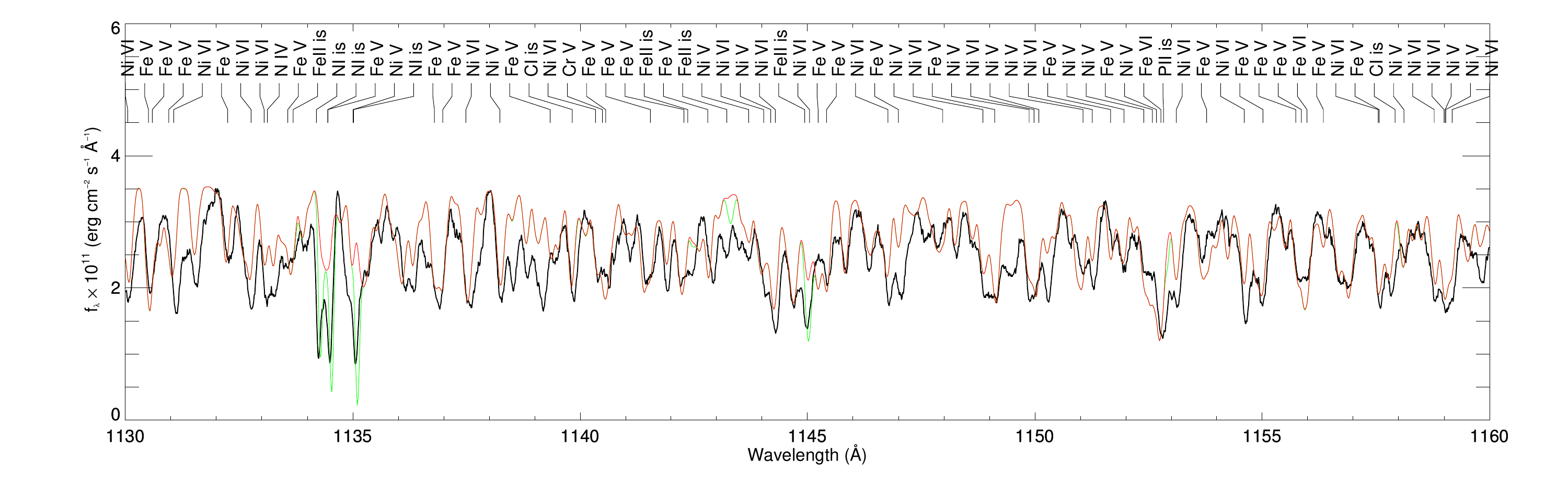}}
\caption{-- Continued. }
\end{center}
\end{figure*}

\addtocounter{figure}{-1}
\begin{figure*}
\begin{center}
\resizebox{\hsize}{!}{\includegraphics[angle=90]{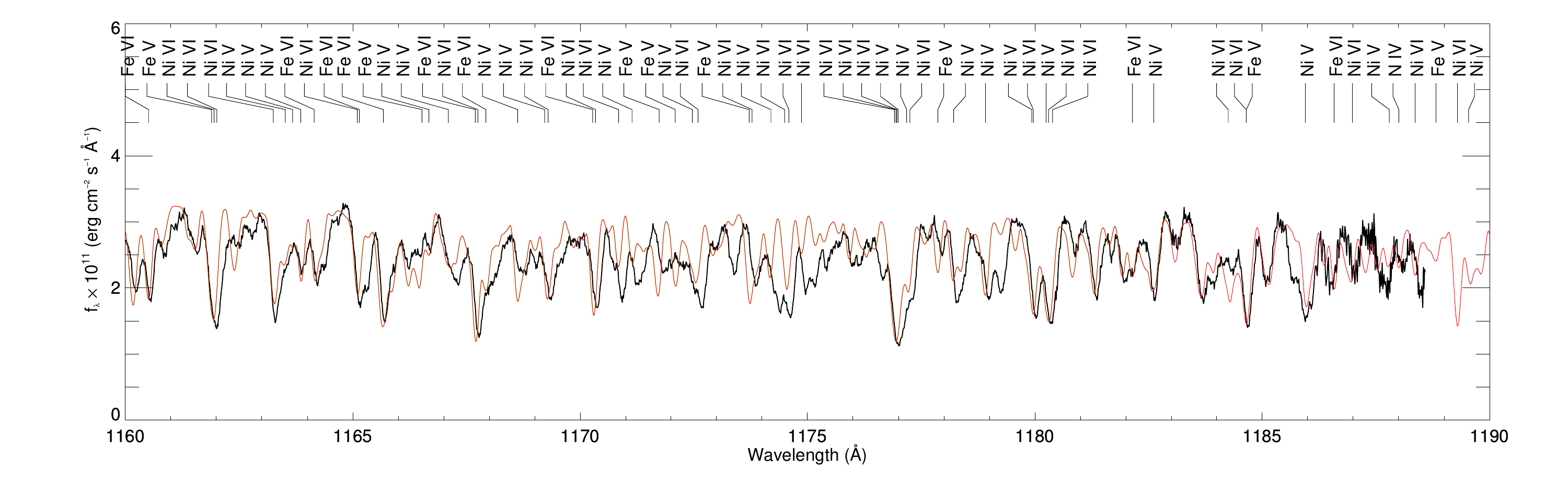}}
\caption{-- Continued. }
\end{center}
\end{figure*}

\begin{figure*}
\begin{center}
\includegraphics[width=1\linewidth]{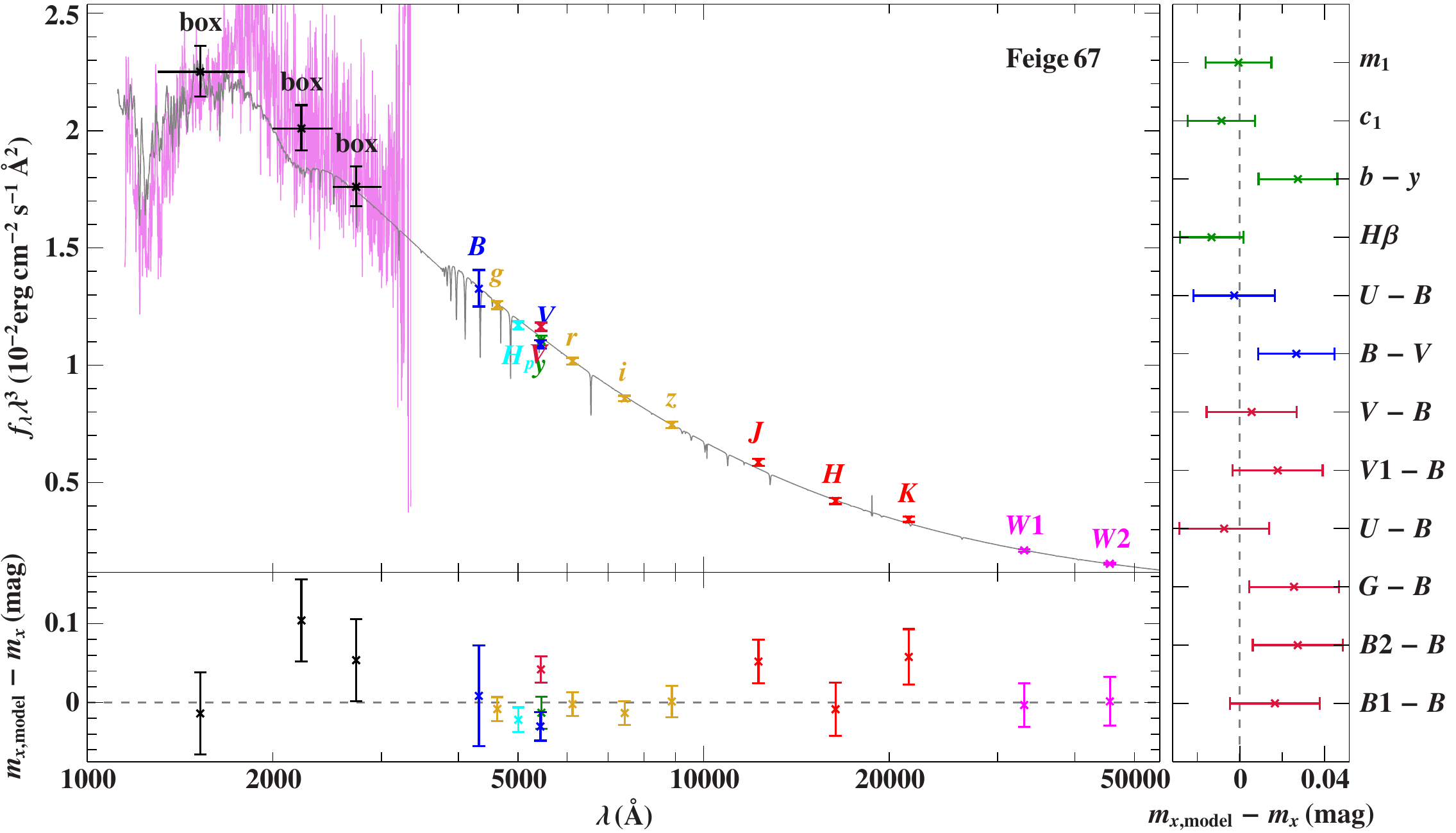}
\caption{Comparison of our best fitting synthetic and observed photometry for 
the single stars \feiget, \agk, and \lsii.
The different photometric systems have the following colour code: Johnson-Cousins 
(blue), Strömgren (green), Tycho (brown), Hipparcos (cyan), SDSS (gold), 2MASS 
(red), WISE (magenta), and Geneva (crimson).
Theoretical reddened spectra are shown in grey and the \textit{IUE} data in 
pink.}
\label{photomadd}
\end{center}
\end{figure*}

\addtocounter{figure}{-1}
\begin{figure*}
\begin{center}
\includegraphics[width=1\linewidth]{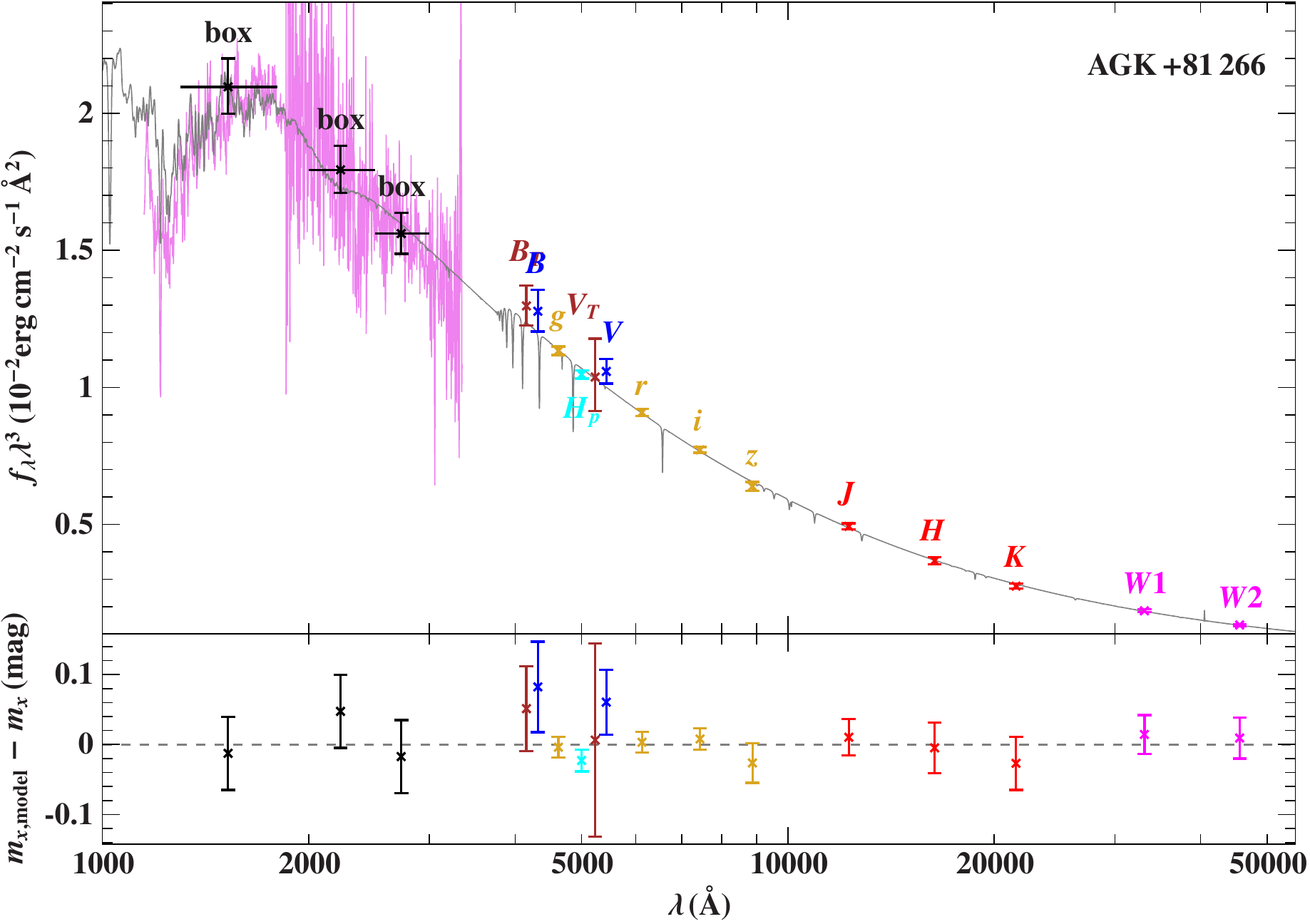}
\includegraphics[width=1\linewidth]{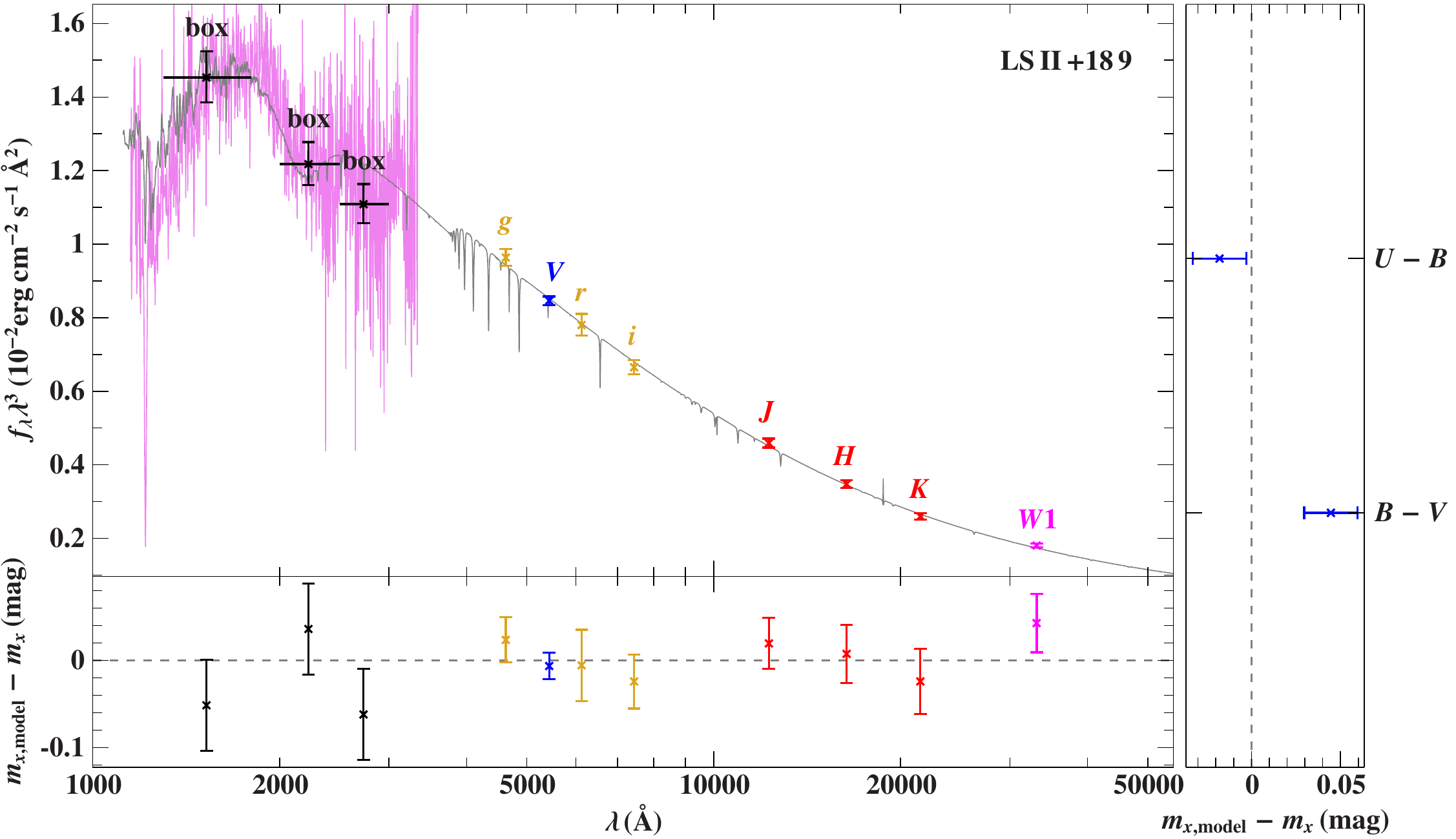}
\caption{-- Continued.}
\end{center}
\end{figure*}

\begin{table}[h]
\caption{References for the photometric data catalogues.}
\label{photocat}
\begin{tabular}{ll}
\hline
\hline
 Photometric system & Reference   \\
\hline
Johnson-Cousins & \citet{mor78} \\
                & \citet{mer97} \\
                & \citet{apass} \\
Strömgren &  \citet{paun15} \\
Tycho & \citet{hog00}\\
Hipparcos & \citet{hip07}  \\
SDSS &  \citet{sdssdr9} \\
      & \citet{apass}  \\
2MASS & \citet{apass} \\
WISE &   \citet{cutri12}   \\
Geneva & \citet{geneva} \\
\hline
\end{tabular}
\end{table}

\end{appendix}

\end{document}